\documentclass[11pt]{cernrep}
\usepackage{epsfig}
\usepackage{graphicx}
\usepackage{cite,./mcite}
\begin{document}
\title{Uncertainties on $W$ and $Z$ production at the LHC \\
HERA - LHC Workshop Proceedings }
\author{Alessandro Tricoli,  Amanda Cooper-Sarkar, Claire Gwenlan}
\institute{Oxford University}
\maketitle
\begin{abstract}
Uncertainties on low-$x$ PDFs are crucial for the standard model benchmark 
processes of $W$ and $Z$ production at the LHC. The current level of PDF 
uncertainty is critically reviewed and the possibility of reducing this 
uncertainty using early LHC data is investigated taking into account realistic 
expectations for measurement accuracy, kinematic cuts and backgrounds.
\end{abstract}

\section{Introduction}
\label{sec:intro}

\begin{figure}[tbp]
\vspace{-2.0cm} 
\centerline{
\epsfig{figure=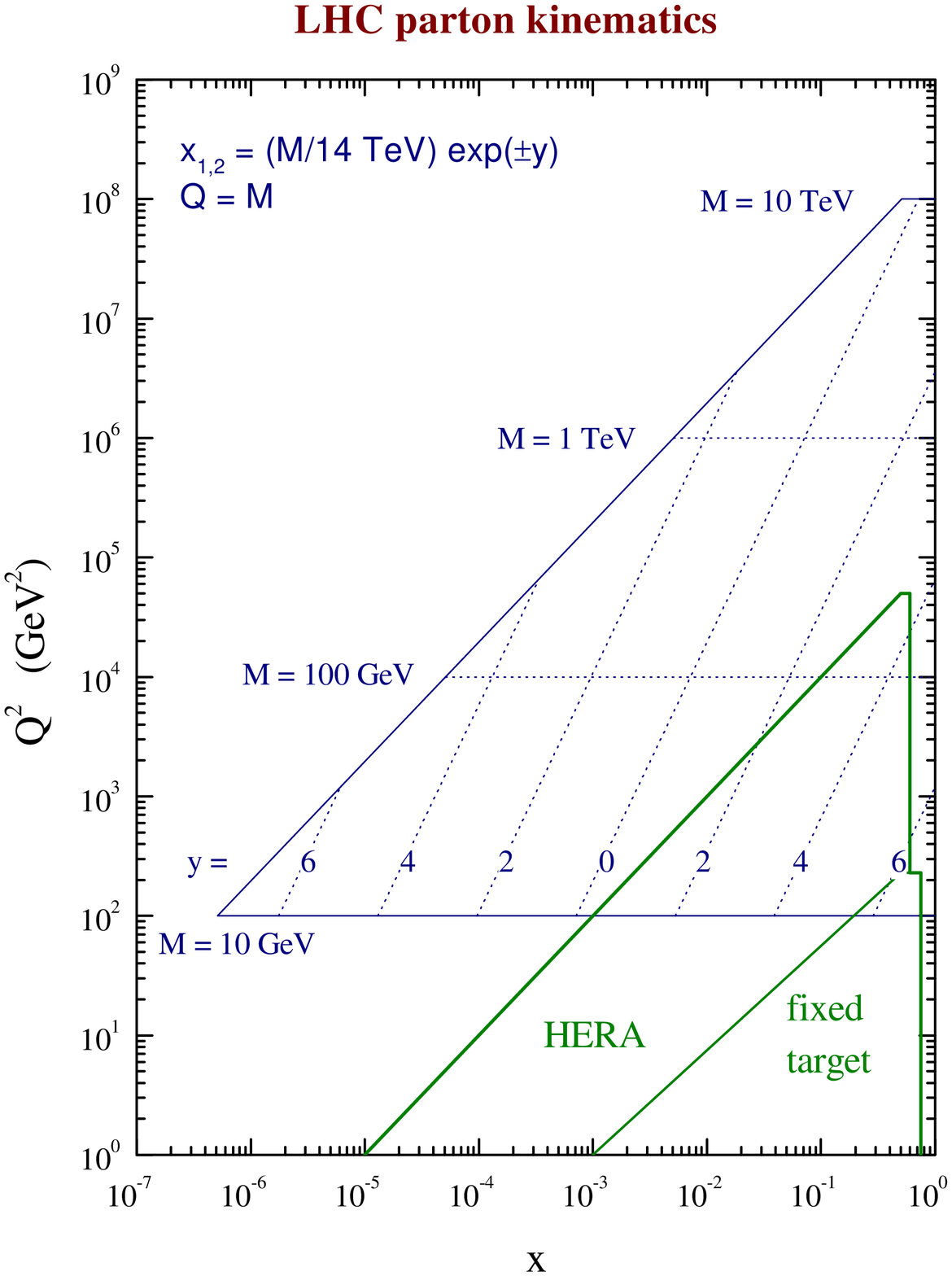,width=0.5\textwidth}
\epsfig{figure=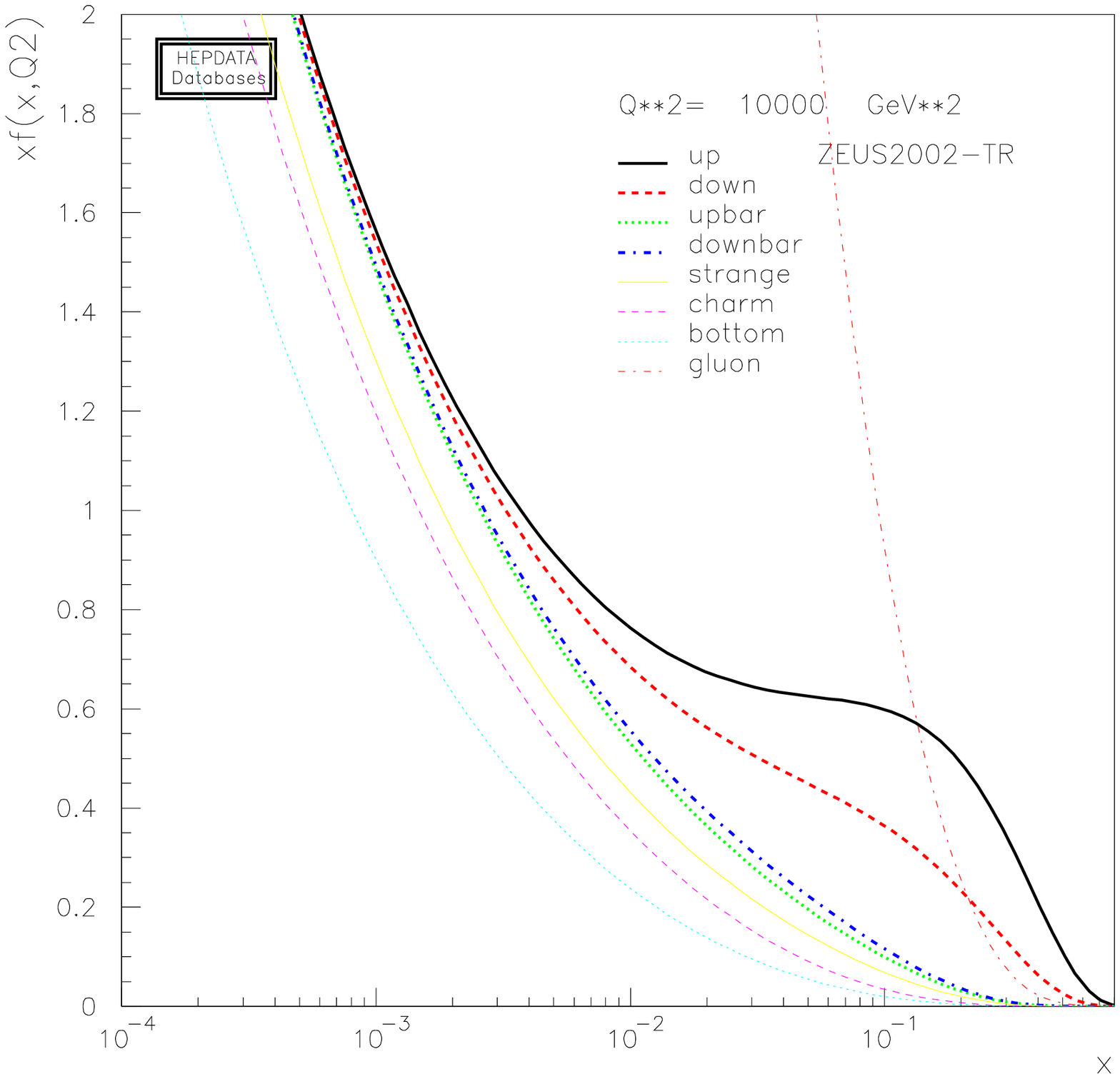,width=0.5\textwidth}}
\caption {Left plot: The LHC kinematic plane (thanks to James Stirling).
Right plot: PDF distributions at $Q^2 = 10,000$~GeV$^2$.}
\label{fig:kin/pdfs}
\end{figure}

 At leading order (LO), $W$ and $Z$ production occur by the process, $q \bar{q} \rightarrow W/Z$, 
and the momentum fractions of the partons 
participating in this subprocess are given by, $x_{1,2} = \frac{M}{\surd{s}} exp (\pm y)$, where 
$M$ is the centre of mass energy of the subprocess, $M = M_W$ or $M_Z$, $\surd{s}$ is the centre of
 mass energy of the reaction  ($\surd{s}=14$ TeV at the LHC) and 
$y = \frac{1}{2} ln{\frac{(E+pl)}{(E-pl)}}$ gives the parton rapidity. 
The kinematic plane for LHC parton kinematics is shown in Fig.~\ref{fig:kin/pdfs}. 
Thus, at central rapidity, the participating partons have small momentum fractions, $x \sim 0.005$.
Moving away from central rapidity sends one parton to lower $x$ and one 
to higher $x$, but over the measurable rapidity range, $|y| < 2.5$, 
$x$ values remain in the range, 
$10^{-4} < x < 0.1$. Thus, in contrast to the situation at the Tevatron, valence quarks are not 
involved, the scattering is happening between sea quarks. Furthermore, the high 
scale of the process $Q^2 = M^2 \sim 10,000$~GeV$^2$ ensures that the gluon is the dominant 
parton, see Fig.~\ref{fig:kin/pdfs}, so that these sea quarks have mostly 
been generated by the flavour blind $g \to q \bar{q}$ splitting process. Thus the precision of 
our knowledge of $W$ and $Z$ cross-sections at the LHC is crucially dependent on the uncertainty on 
the momentum distribution of the gluon. 

\begin{figure}[tbp] 
\centerline{
\epsfig{figure=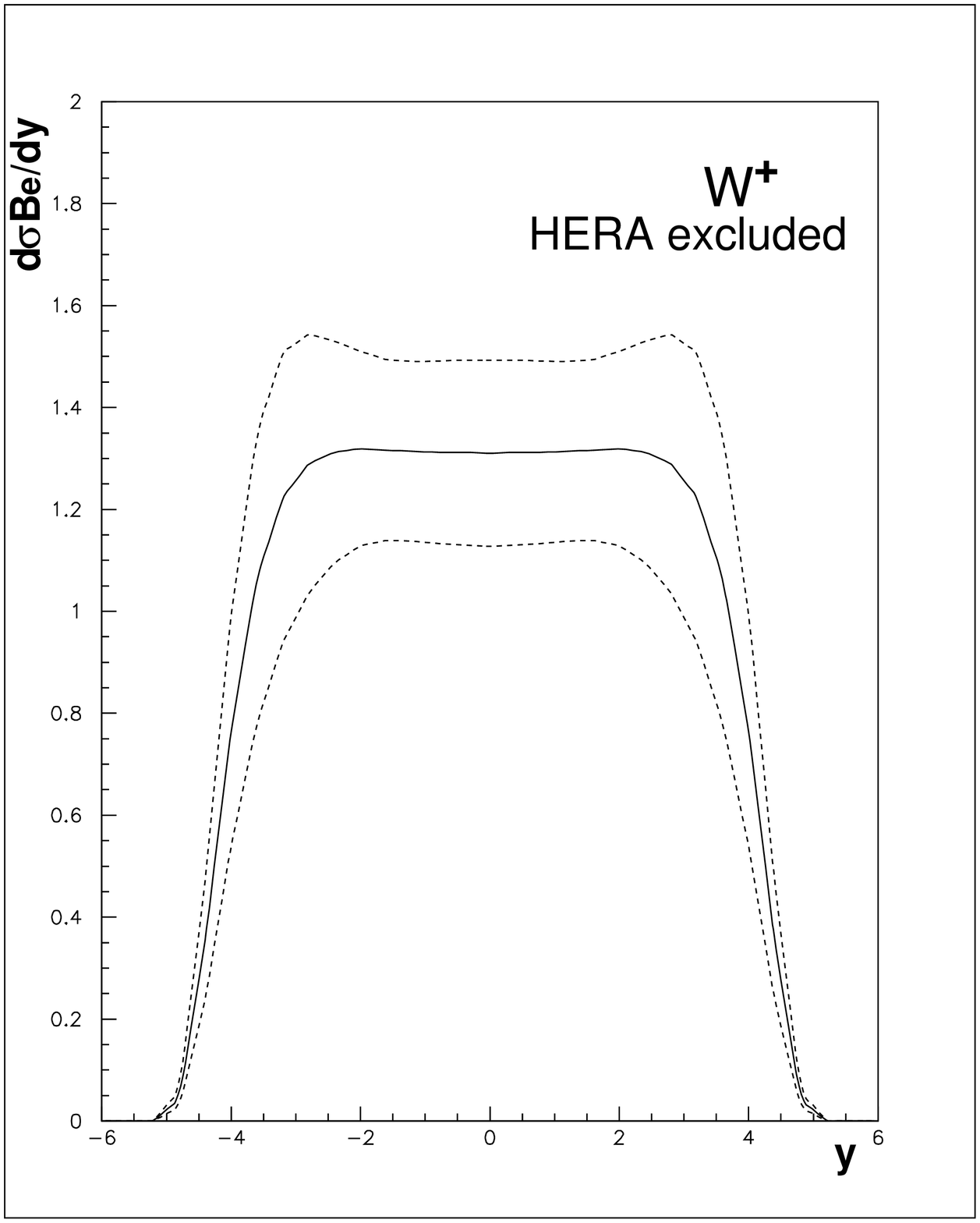,width=0.3\textwidth,height=5cm}
\epsfig{figure=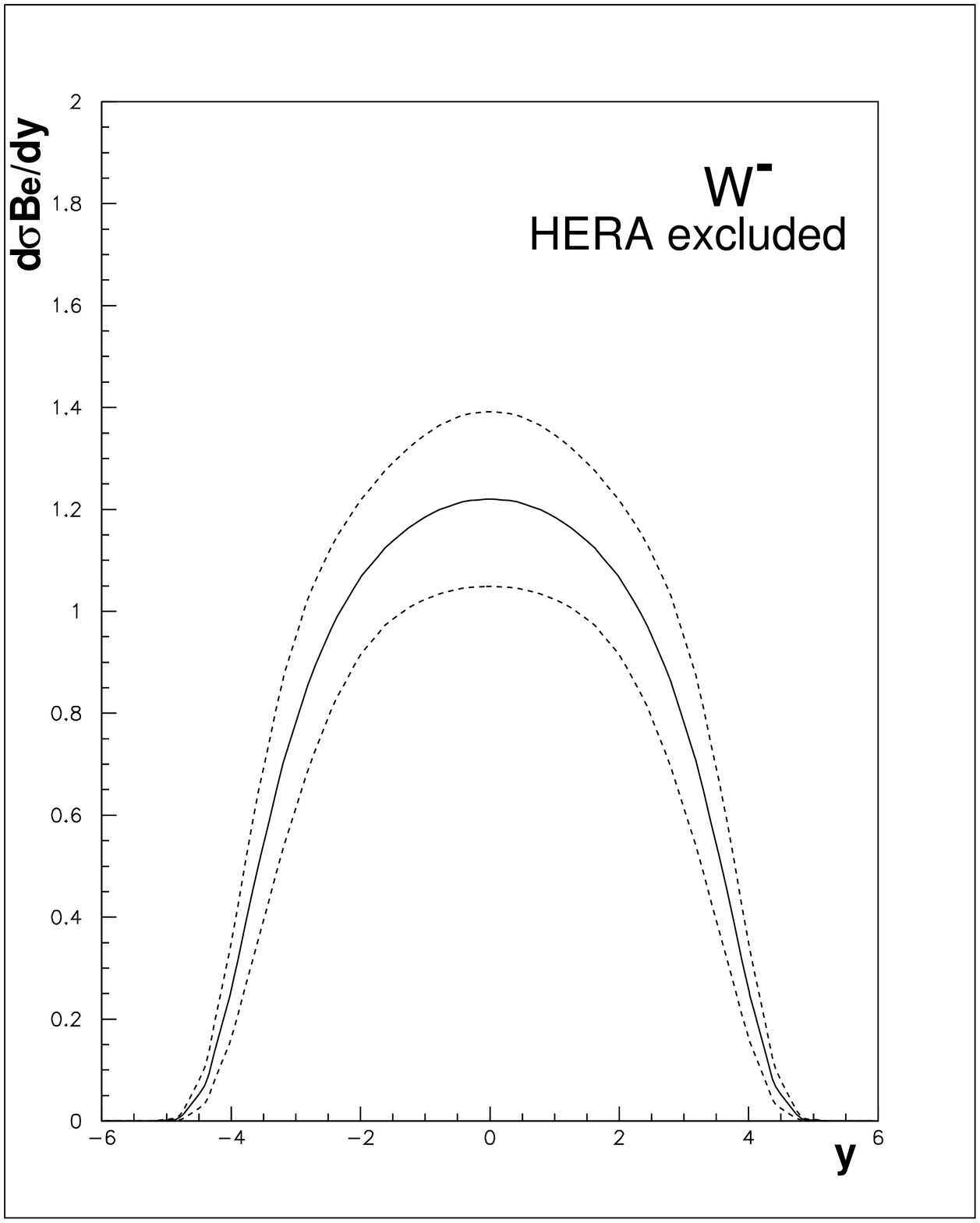,width=0.3\textwidth,height=5cm}
\epsfig{figure=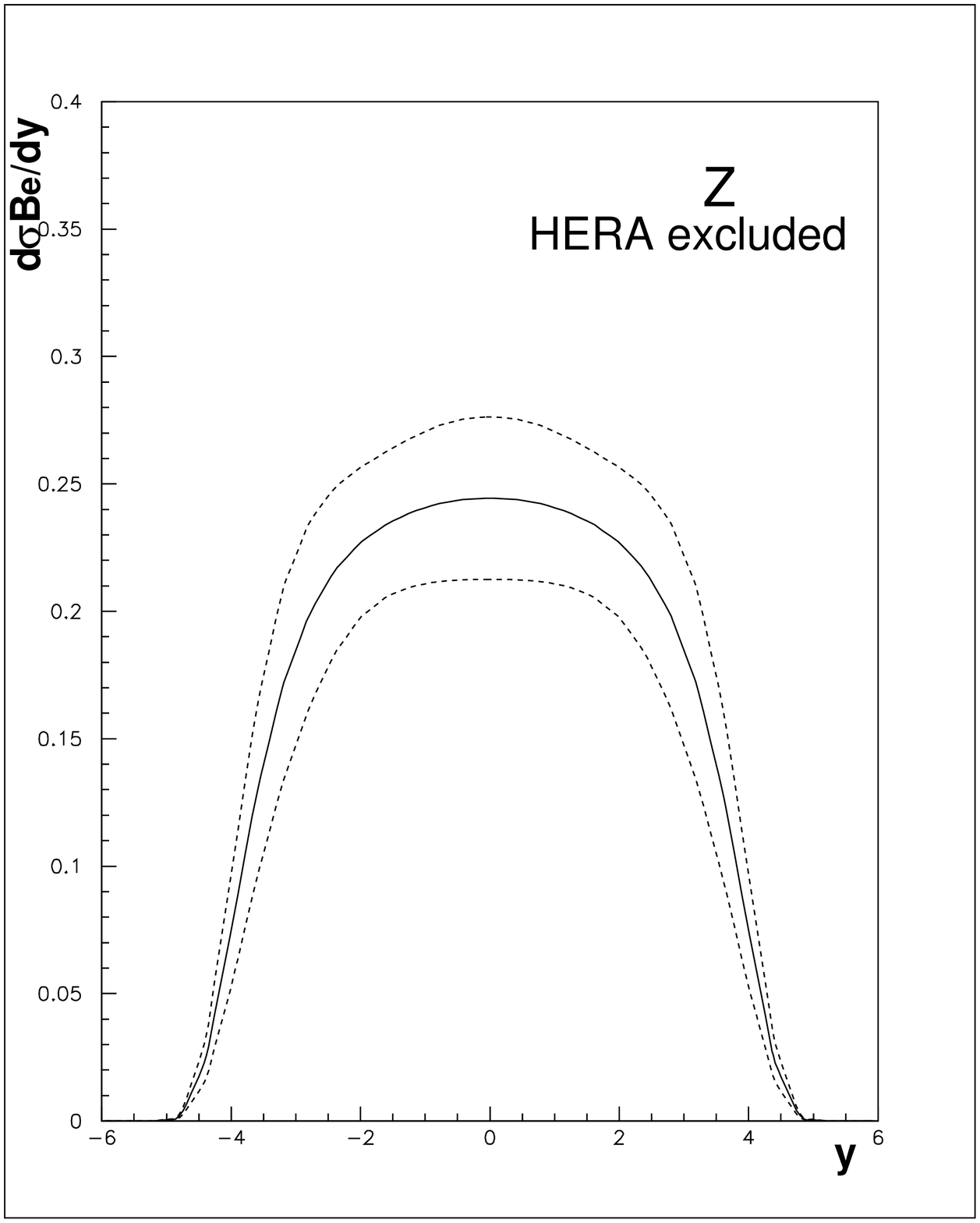,width=0.3\textwidth,height=5cm} 
}
\centerline{
\epsfig{figure=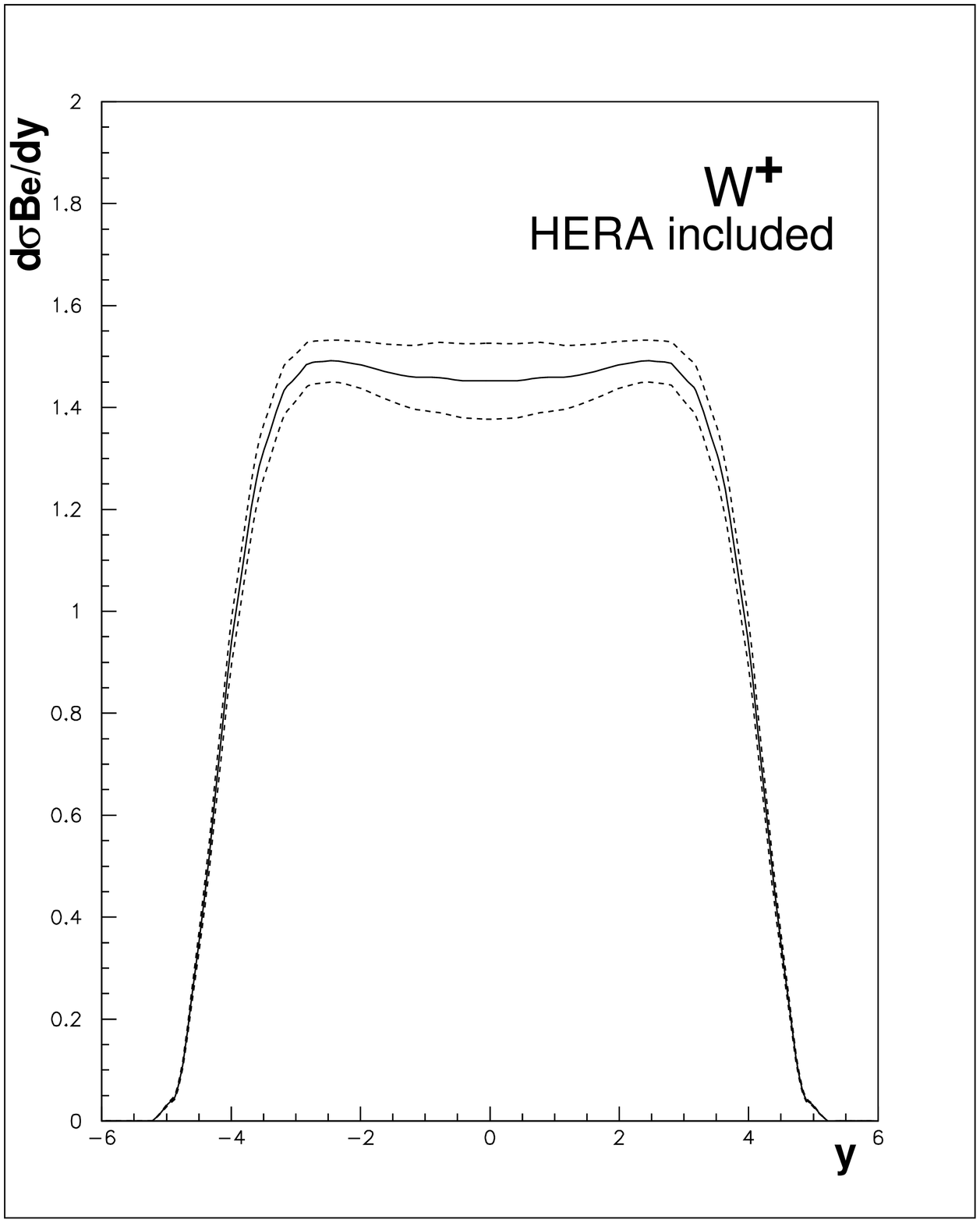,width=0.3\textwidth,height=5cm}
\epsfig{figure=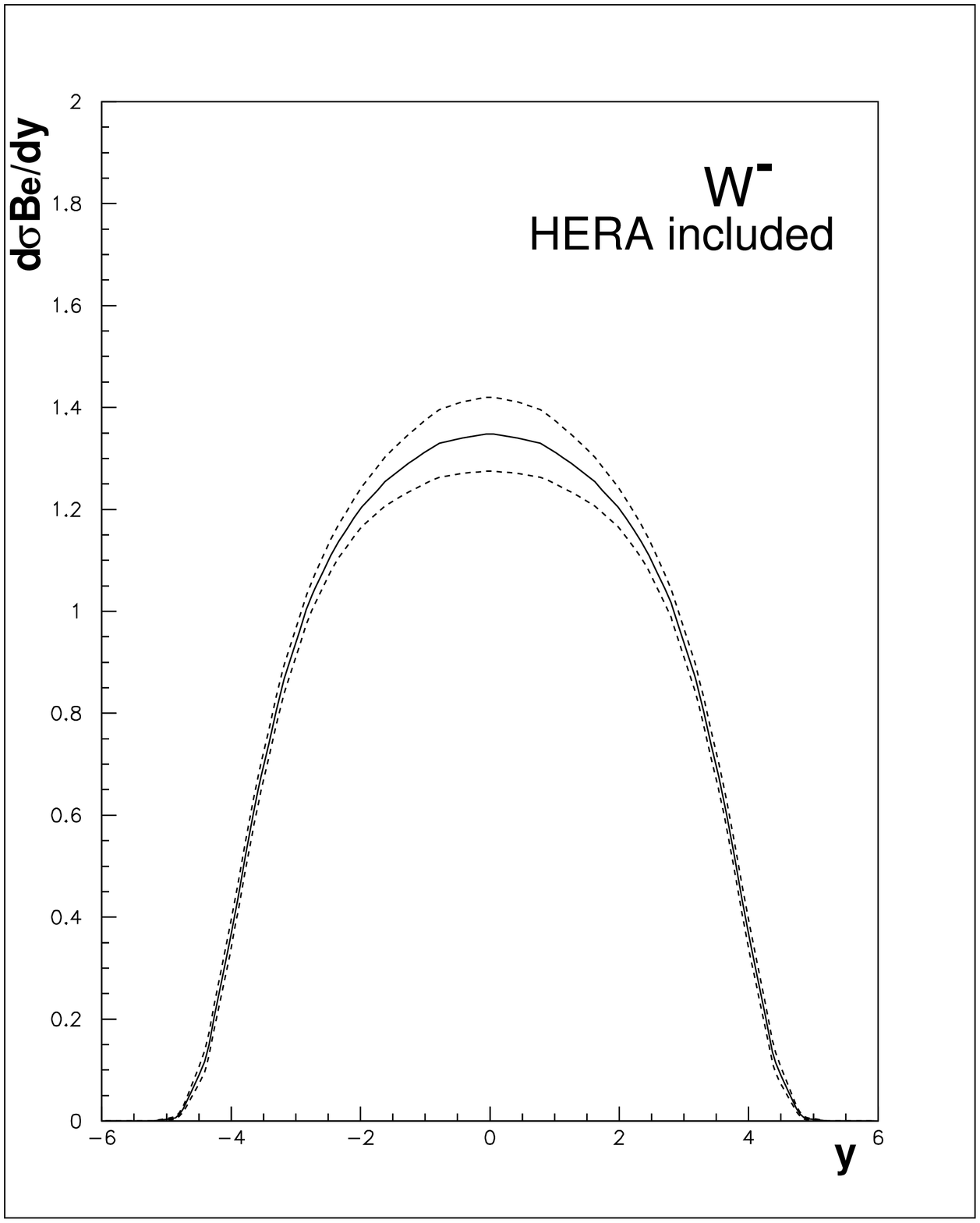,width=0.3\textwidth,height=5cm}
\epsfig{figure=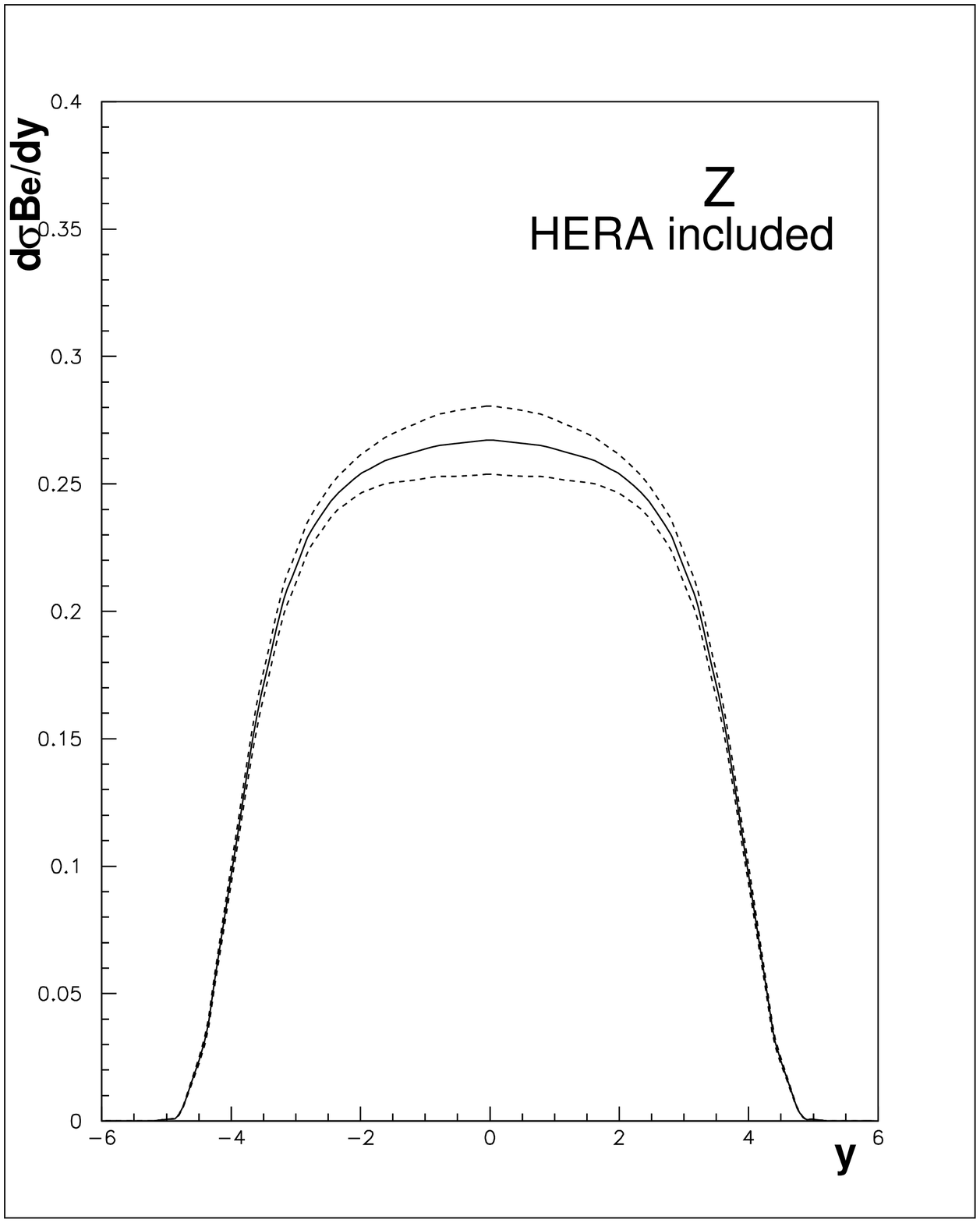,width=0.3\textwidth,height=5cm}
}
\caption {LHC $W^+,W^-,Z$ rapidity distributions and their PDF uncertainties 
(the full line shows the central value and the dashed lines show the spread of 
the uncertainty): 
Top Row: from the ZEUS-S global PDF analysis
not including HERA data; left plot $W^+$; middle plot $W^-$; right plot $Z$: Bottom Row: from the ZEUS-S
global PDF analysis including HERA data; left plot $W^+$; middle plot $W^-$; right plot $Z$}
\label{fig:WZrapFTZS13}
\end{figure}

\begin{figure}[tbp] 
\vspace{-1.0cm}
\centerline{
\epsfig{figure=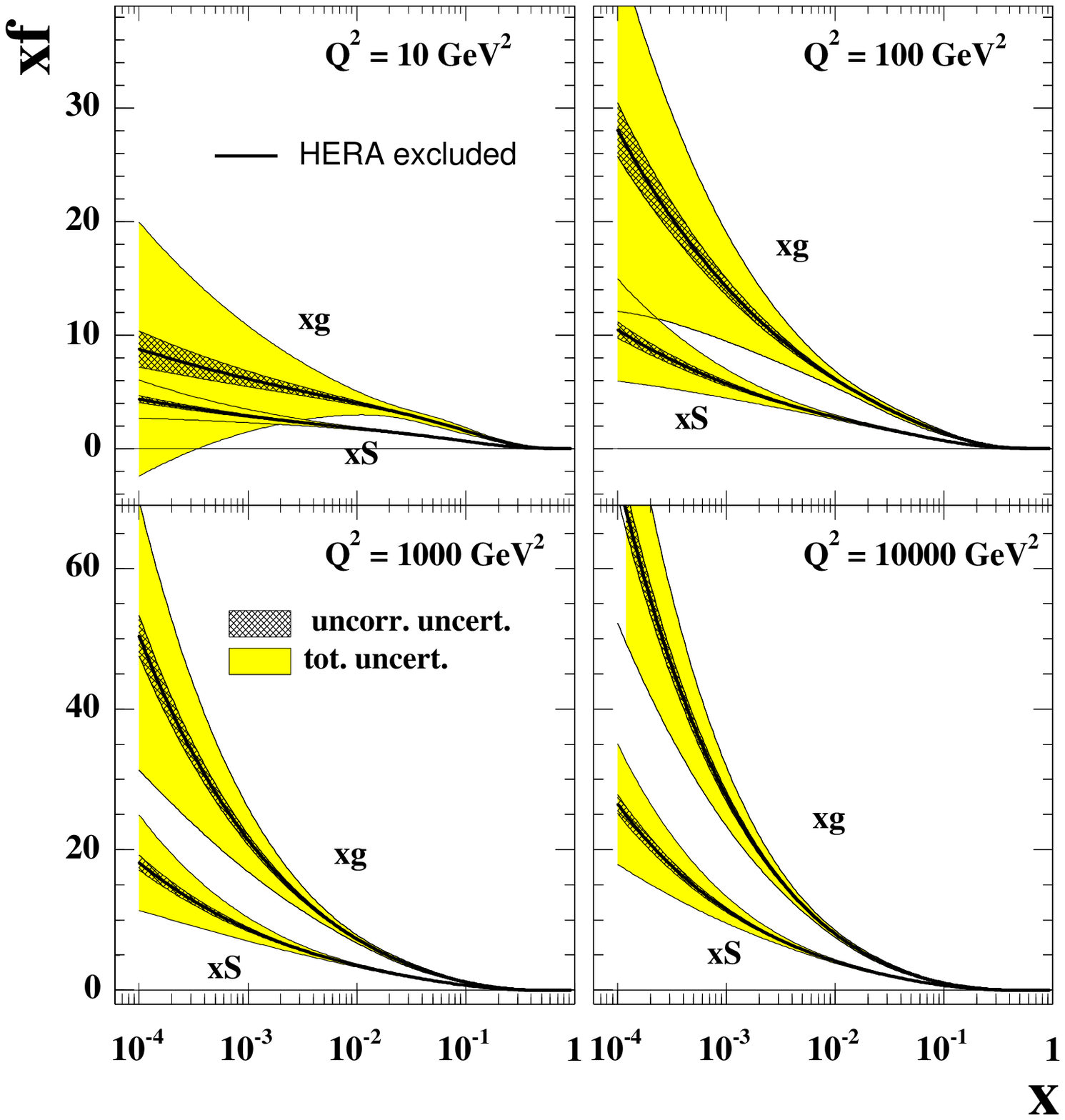,width=0.5\textwidth}
\epsfig{figure=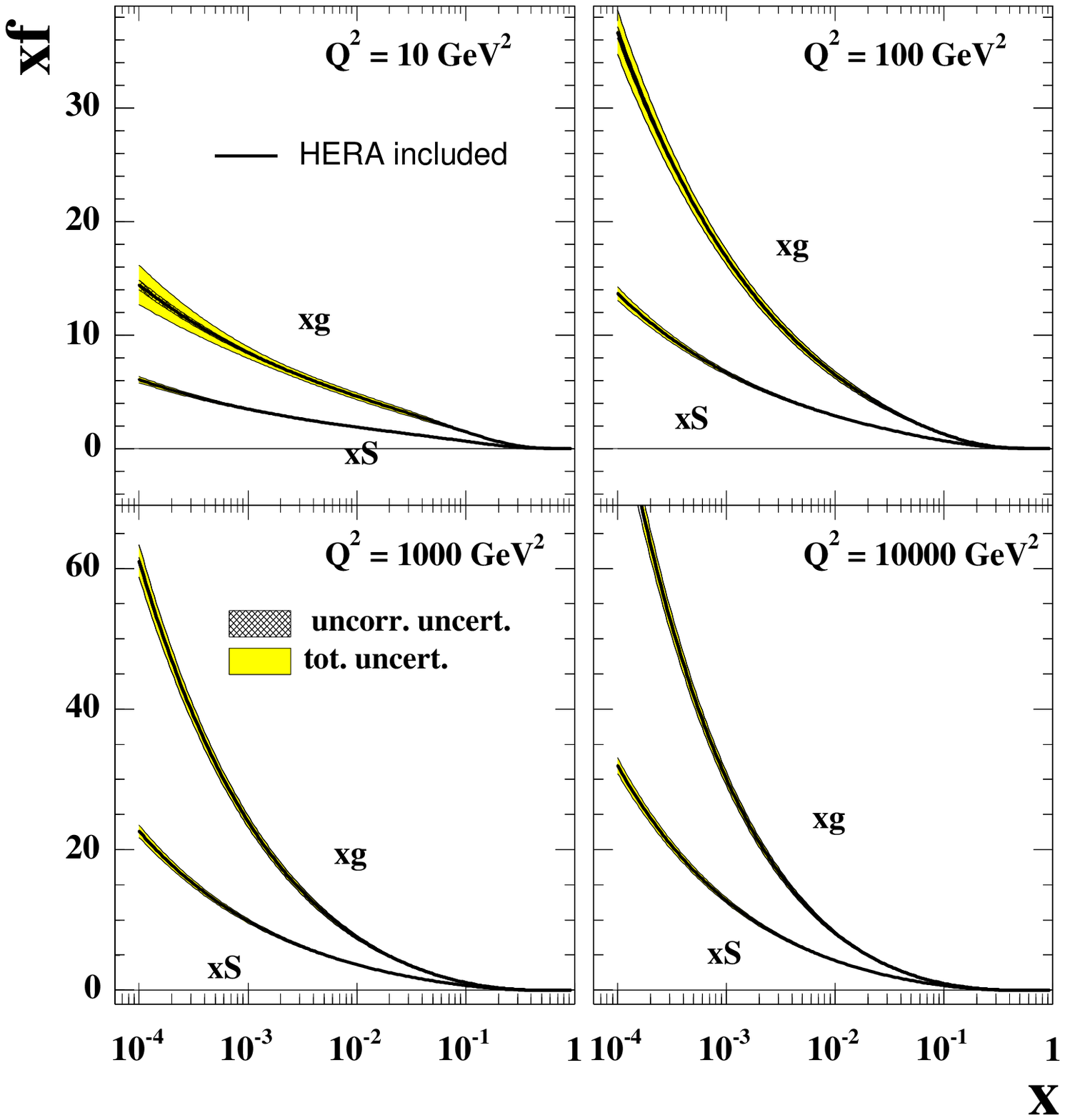,width=0.5\textwidth}}
\caption {Sea ($xS$) and gluon ($xg$) PDFs at various $Q^2$: left plot; 
from the ZEUS-S global PDF analysis
 not including HERA data; right plot: from the ZEUS-S global PDF analysis 
including HERA data. The inner cross-hatched error bands show the statistical 
and uncorrelated systematic uncertainty, the outer error bands show the total 
uncertainty including experimental correlated systematic uncertainties, 
normalisations and model uncertainty.}
\label{fig:pre/postPDFs}
\end{figure}

HERA data have dramatically improved our knowledge of the gluon, as illustrated in 
Fig.~\ref{fig:WZrapFTZS13}, 
which shows $W$ and $Z$ rapidity spectra predicted from a 
global PDF fit which does not include the HERA data, compared to a fit including HERA data. 
The latter fit is the ZEUS-S global fit~\cite{zeus-s}, whereas the former is 
a fit using the same fitting analysis but leaving out the ZEUS data. The full 
PDF uncertainties for both fits are calculated from the error PDF sets of the ZEUS-S analysis
using LHAPDF~\cite{LHAPDF} (see the contribution of M.Whalley to these proceedings). 
The predictions for the $W/Z$ cross-sections, decaying to the lepton decay mode, are summarised in 
Table~\ref{tab:datsum}.
\begin{table}[t]
\centerline{\small
\begin{tabular}{llllcccc}\\
 \hline
PDF Set  & $\sigma(W^+).B(W^+ \rightarrow l^+\nu_l)$ & $\sigma(W^-).B(W^- \rightarrow l^-\bar{\nu}_l)$ & 
$\sigma(Z).B(Z \rightarrow l^+ l^-)$\\
 \hline
 ZEUS-S no HERA  & $10.63 \pm 1.73 $~nb & $7.80 \pm 1.18 $~nb & $1.69 \pm 0.23$~nb \\
 ZEUS-S  & $12.07 \pm 0.41 $~nb & $8.76 \pm 0.30 $~nb & $1.89 \pm 0.06$~nb\\
 CTEQ6.1 & $11.66 \pm 0.56 $~nb & $8.58 \pm 0.43 $~nb & $1.92 \pm 0.08$~nb\\
 MRST01 & $11.72 \pm 0.23 $~nb & $8.72 \pm 0.16 $~nb & $1.96 \pm 0.03$~nb\\
 \hline\\
\end{tabular}}
\caption{LHC $W/Z$ cross-sections for decay via the lepton mode, for various PDFs}
\label{tab:datsum}
\end{table}
The uncertainties 
in the predictions for these cross-sections have decreased from $\sim 16\%$ pre-HERA to $\sim 3.5\%$ 
post-HERA. The reason for this can be seen clearly in Fig.~\ref{fig:pre/postPDFs}, 
where the sea and gluon 
distributions for the pre- and post-HERA fits are shown for several different $Q^2$ bins, together 
with their uncertainty bands. It is the dramatically increased precision in the low-$x$ gluon PDF,
feeding into increased precision in the low-$x$ sea quarks, 
which has led to the increased precision on the predictions for $W/Z$ production at the LHC. 

\begin{figure}[tbp] 
\centerline{
\epsfig{figure=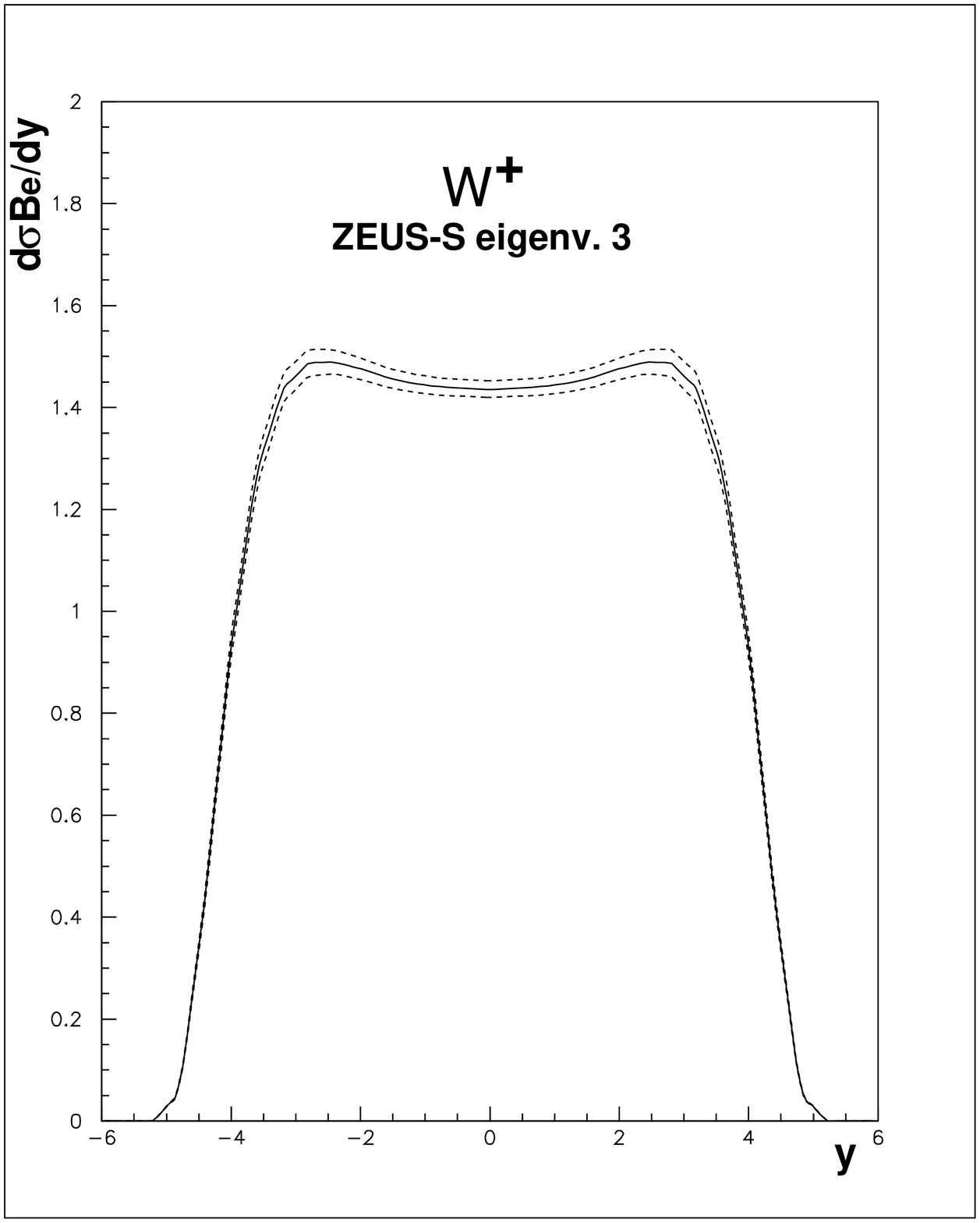,width=0.3\textwidth,height=5cm} 
\epsfig{figure=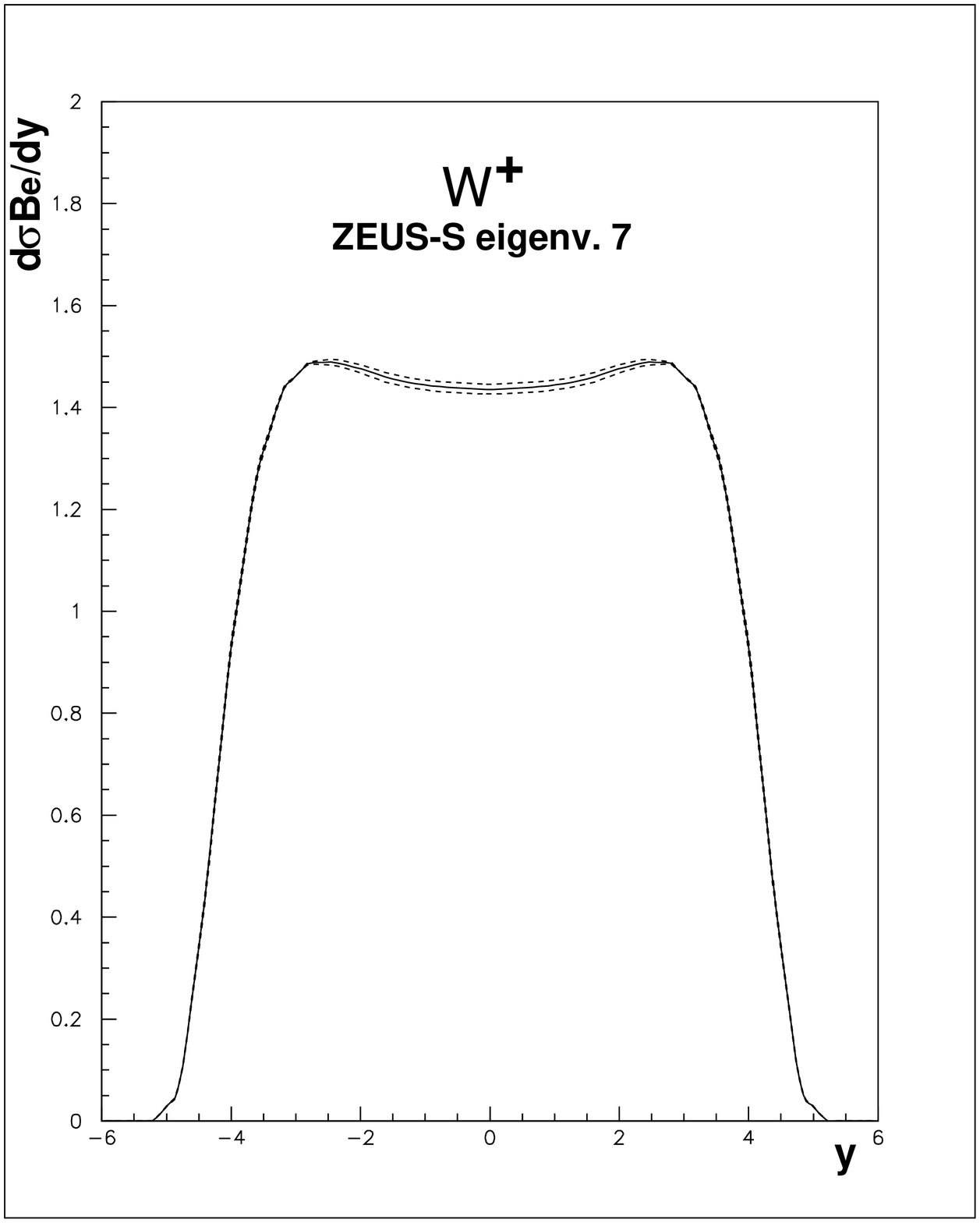,width=0.3\textwidth,height=5cm}
\epsfig{figure=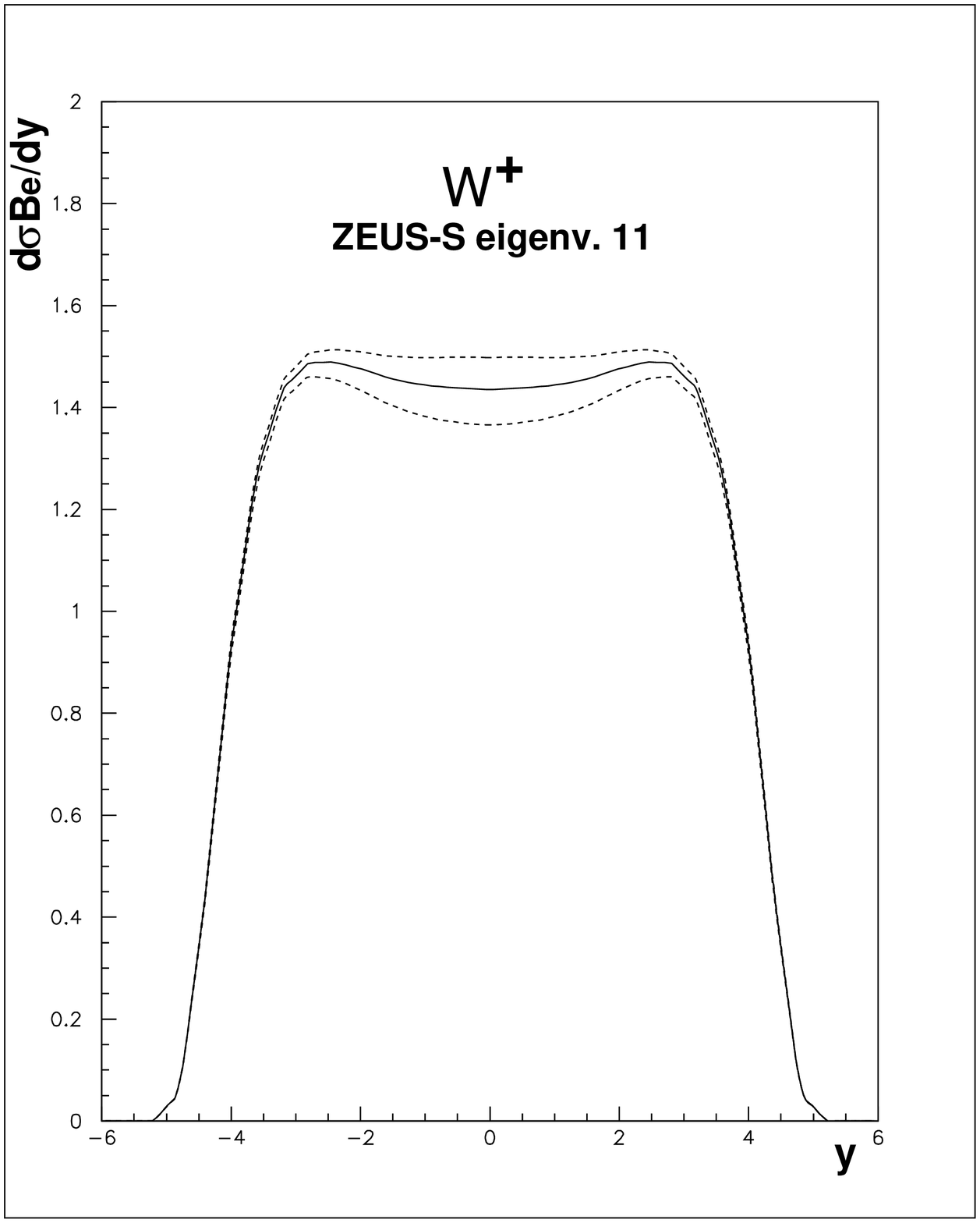,width=0.3\textwidth,height=5cm}
}
\caption {LHC $W^+$ rapidity distributions and their PDF uncertainties due to the eigenvectors 3,7 and 11 of the 
ZEUS-S analysis.}
\label{fig:eigen}
\end{figure}

Further evidence for the conclusion that the uncertainties on the gluon PDF at the input scale ($Q^2_0=7$~GeV$^2$,
 for ZEUS-S) are the major 
contributors to the uncertainty on the $W/Z$ cross-sections at $Q^2= M_W(M_Z)$, comes from decomposing the 
predictions down into their contributing eigenvectors. Fig~\ref{fig:eigen} shows the dominant contributions to the total 
uncertainty from eigenvectors 3, 7, and 11 which are eigenvectors which are
dominated by the parameters which control the low-$x$, mid-$x$ and high-$x$, gluon respectively. 

\begin{figure}[tbp] 
\centerline{
\epsfig{figure=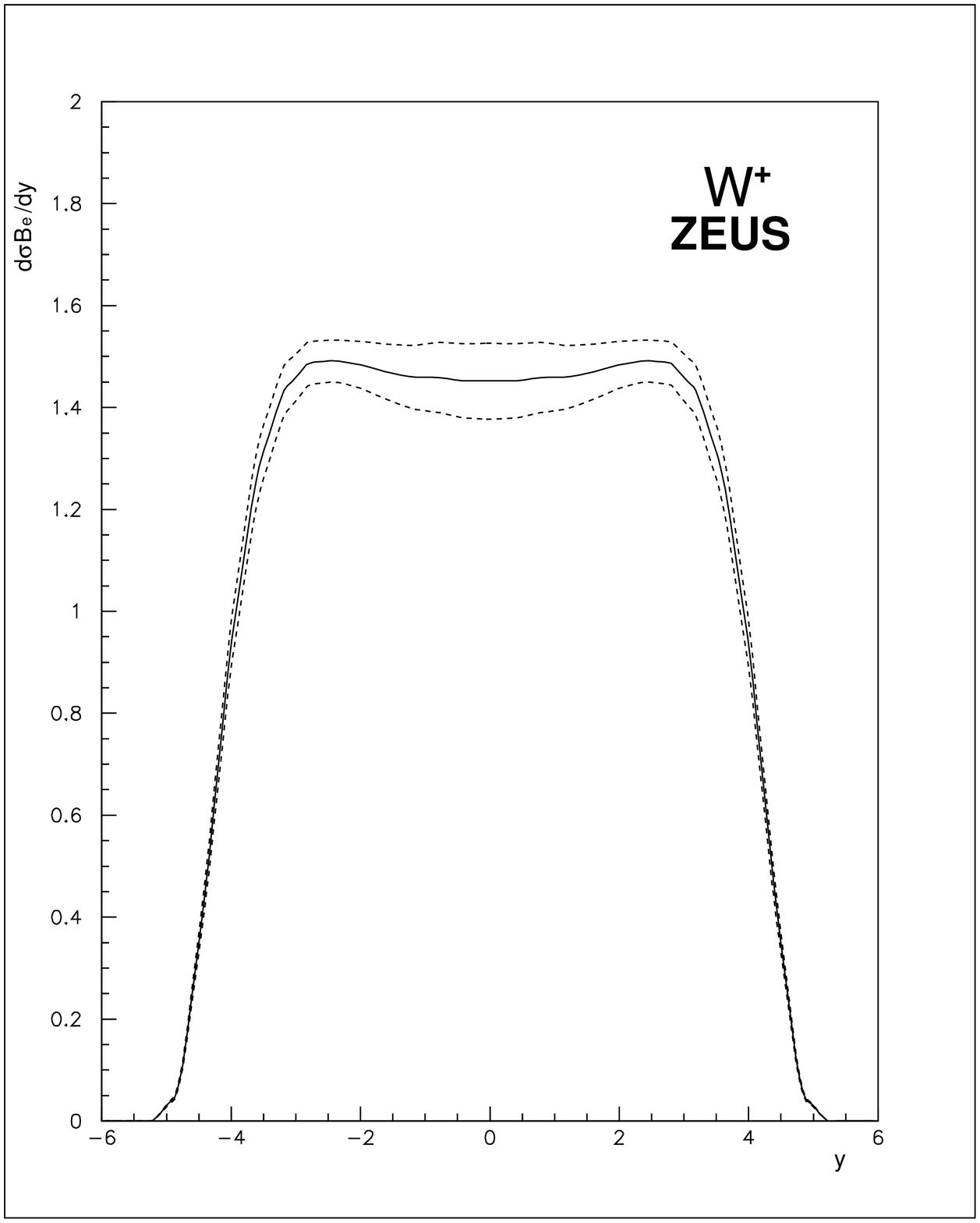,width=0.3\textwidth,height=5cm}
\epsfig{figure=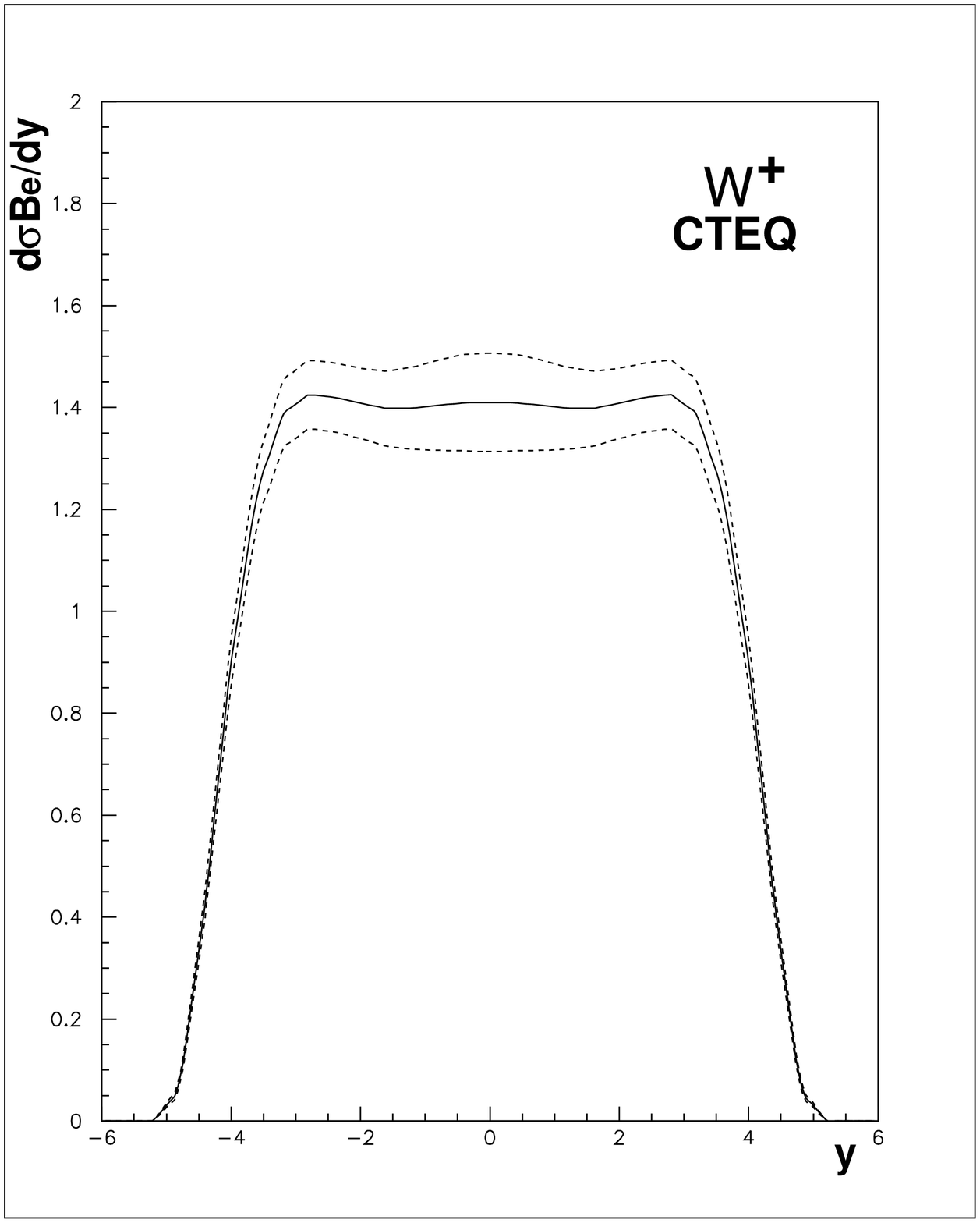,width=0.3\textwidth,height=5cm}
\epsfig{figure=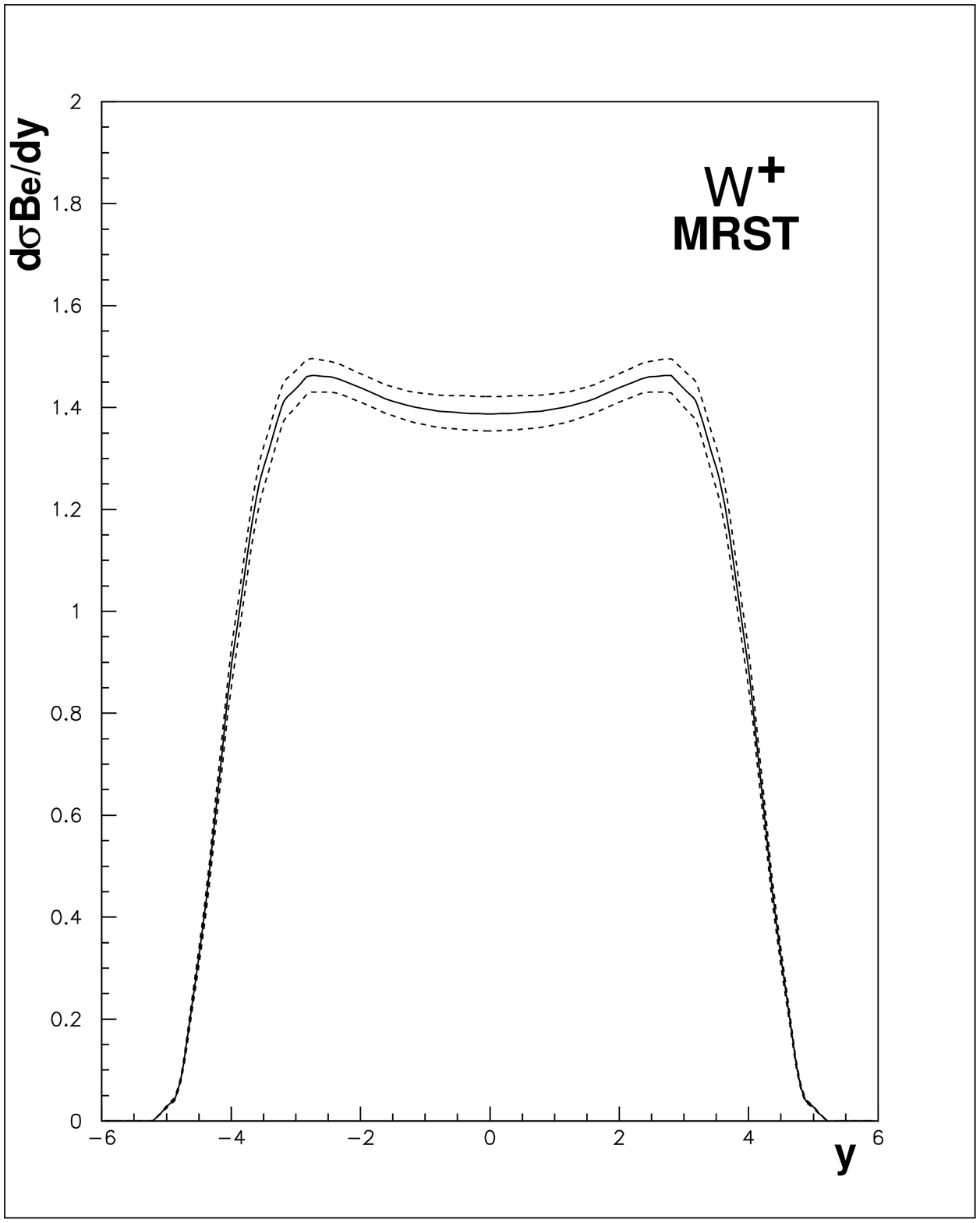,width=0.3\textwidth,height=5cm}
}
\caption {LHC $W^+$ rapidity distributions and their PDF uncertainties:
 left plot, ZEUS-S PDFs; middle plot, CTEQ6.1 PDFs;
right plot: MRST01 PDFs.}
\label{fig:mrstcteq}
\end{figure}

The post-HERA level of precision illustrated in Fig.~\ref{fig:WZrapFTZS13} 
is taken for granted in modern analyses, such that $W/Z$ production have 
been suggested as `standard-candle' processes for luminosity measurement. However, when 
considering the PDF uncertainties on the Standard Model (SM) predictions it is necessary not 
only to consider the uncertainties of a particular PDF analysis, but also to compare PDF 
analyses. Fig.~\ref{fig:mrstcteq} compares the predictions for $W^+$ production for the ZEUS-S PDFs 
with those of 
the CTEQ6.1\cite{cteq} PDFs and the MRST01\cite{mrst} PDFs\footnote{MRST01 PDFs are used because the 
full error analysis is available only for this PDF set.}. 
The corresponding $W^+$ cross-sections, for decay to leptonic mode are 
given in Table~\ref{tab:datsum}.
Comparing the uncertainty at central rapidity, rather 
than the total cross-section, we see that the uncertainty estimates are rather larger: $~5.2\%$ for ZEUS-S; 
$~8.7\%$ 
for CTEQ6.1M and about $~3.6\%$ for MRST01. The difference in the central value between 
ZEUS-S and CTEQ6.1 is $~3.5\%$. Thus the spread in the predictions of the different PDF sets is 
comparable to the uncertainty estimated by the individual analyses. Taking all of these 
analyses together the uncertainty at central rapidity is about $~8\%$. 

\begin{figure}[tbp] 
\vspace{-2.0cm}
\centerline{
\epsfig{figure=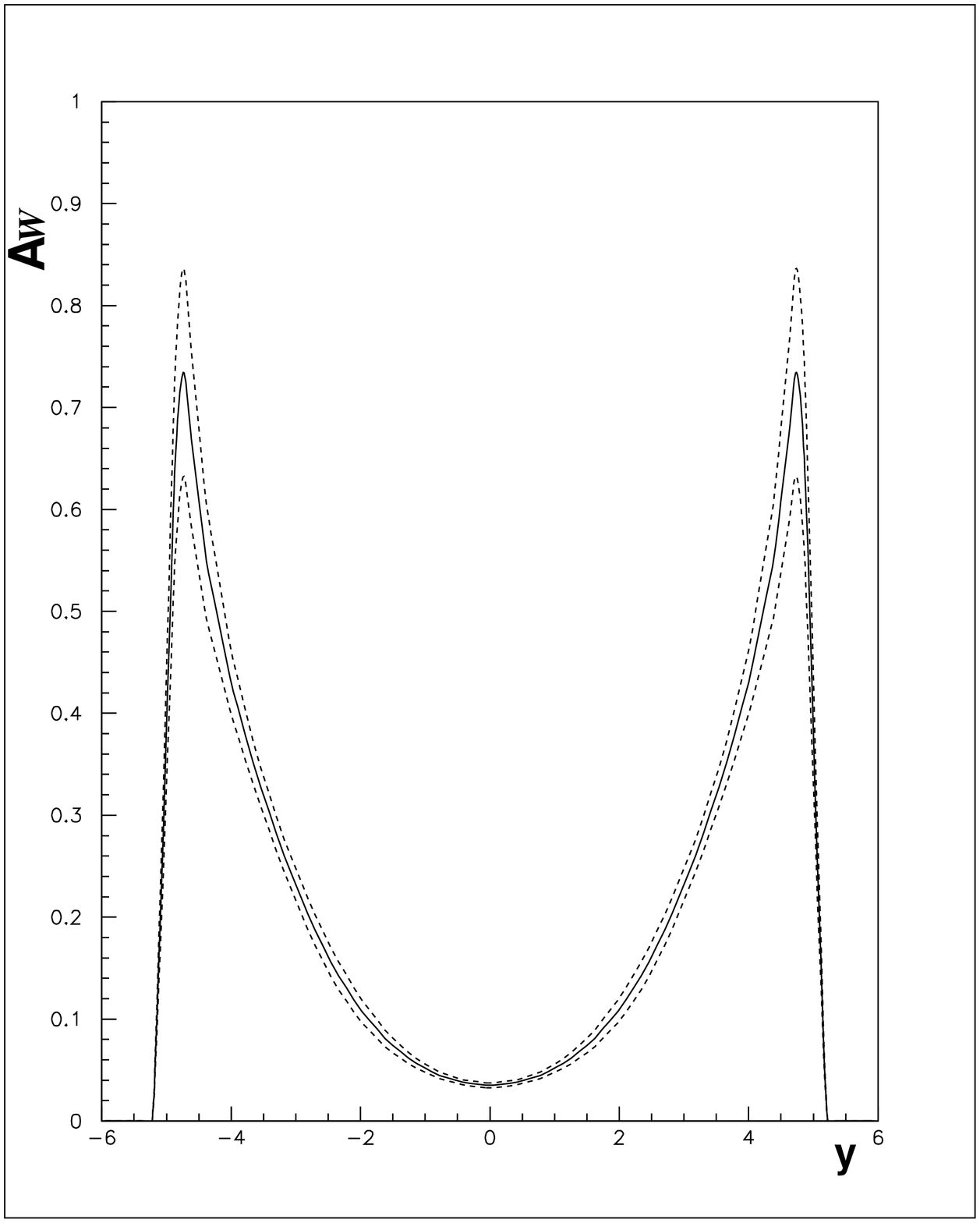,width=0.3\textwidth,height=5cm}
\epsfig{figure=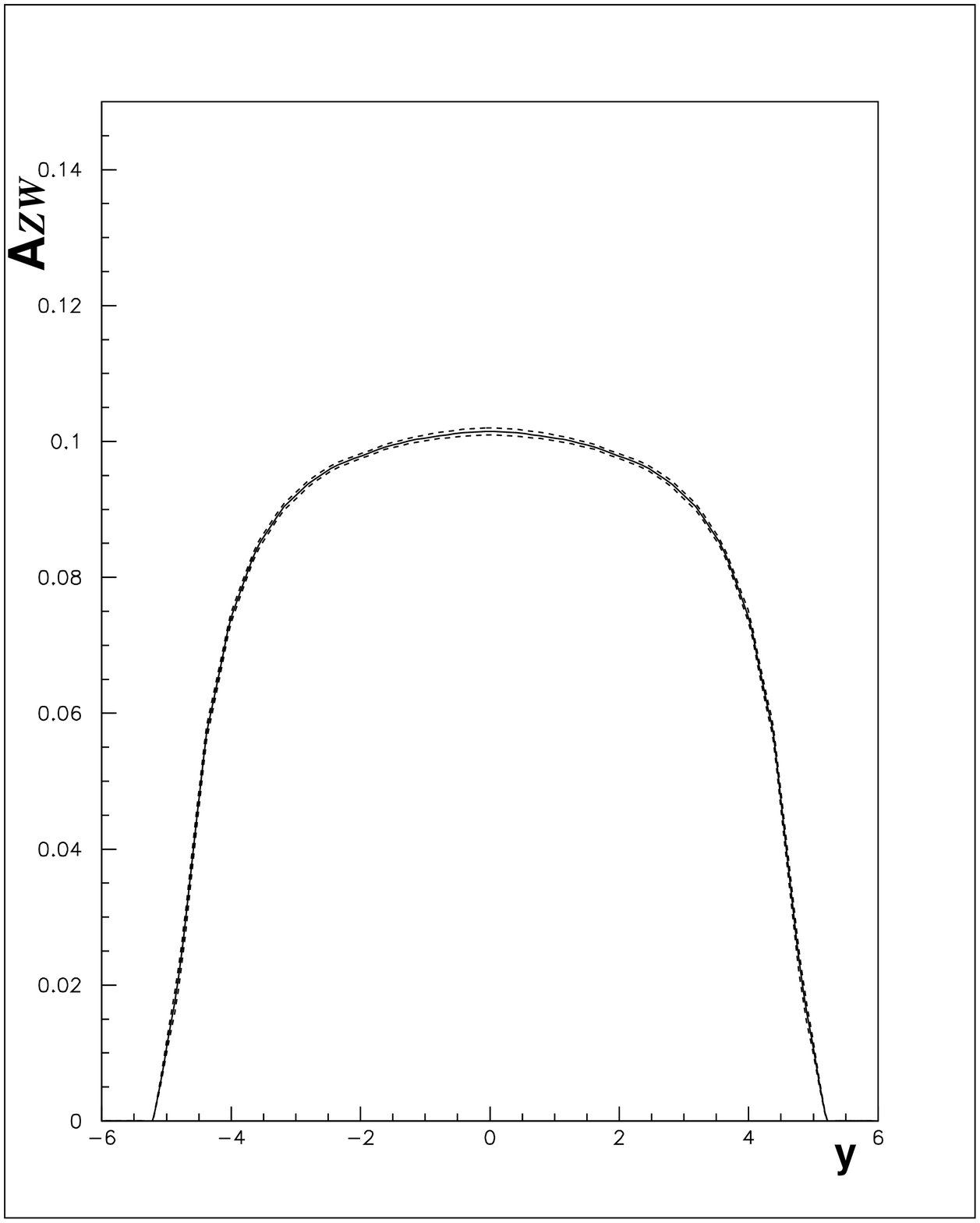,width=0.3\textwidth,height=5cm}
}
\centerline{
\epsfig{figure=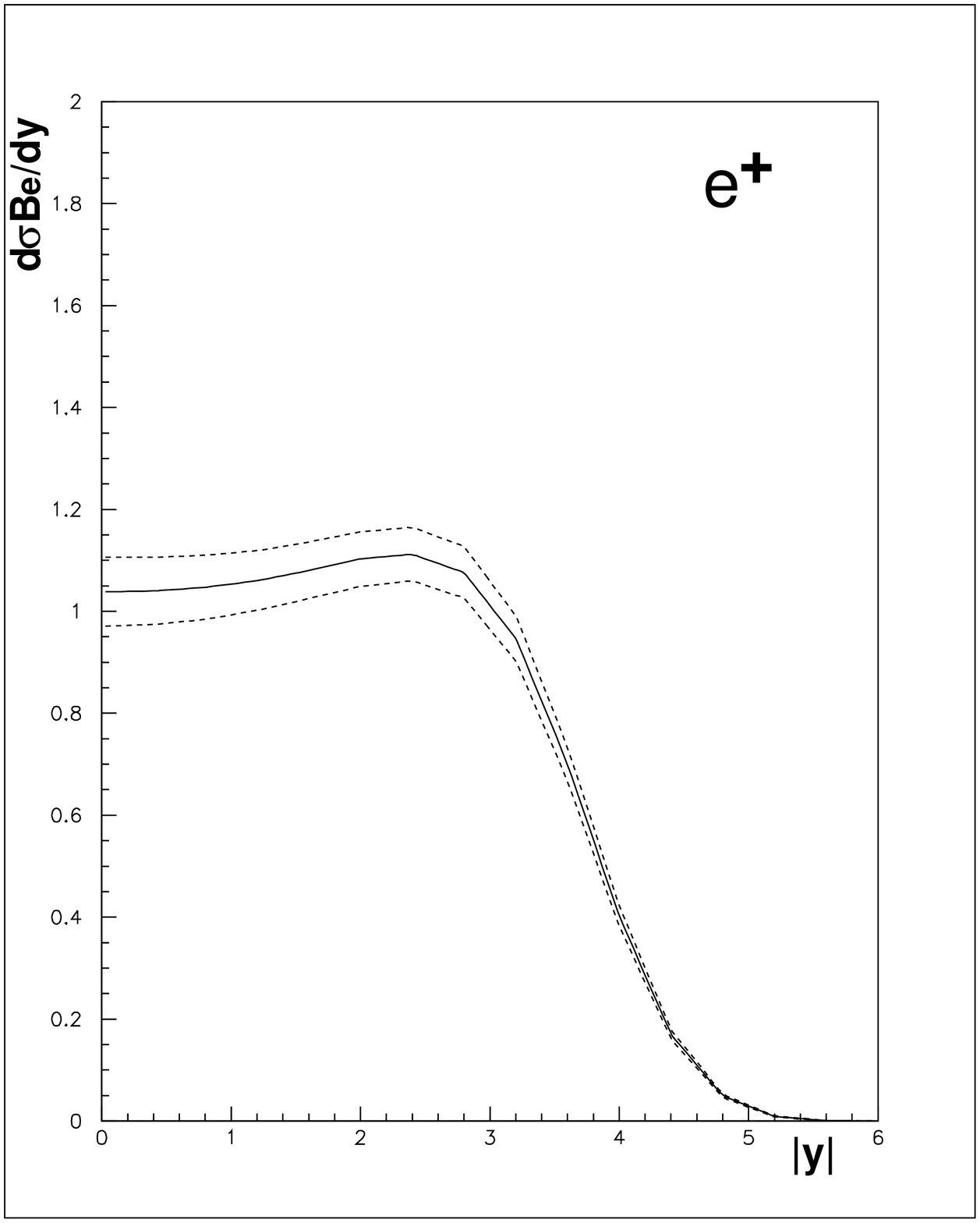,width=0.3\textwidth,height=5cm}
\epsfig{figure=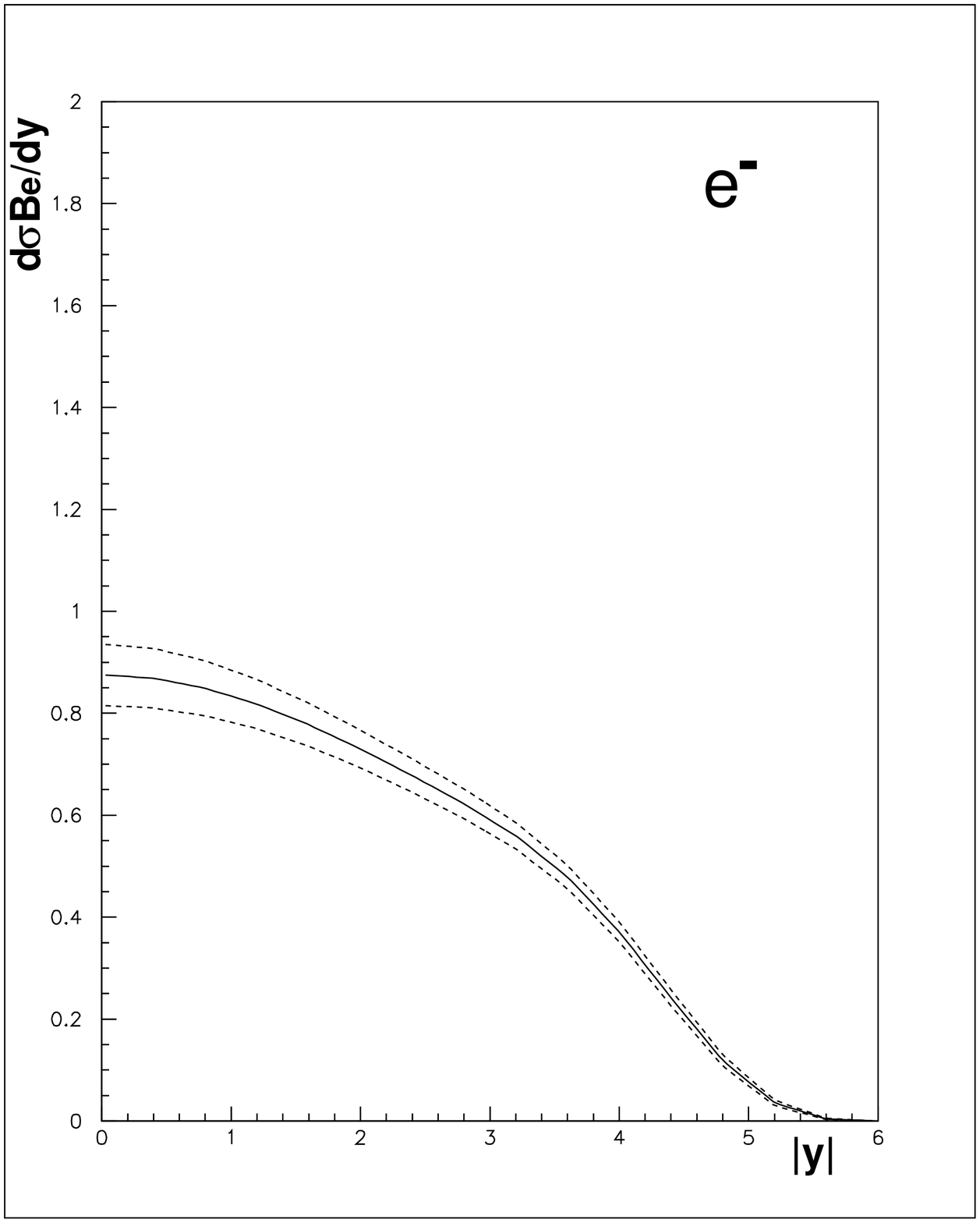,width=0.3\textwidth,height=5cm}
\epsfig{figure=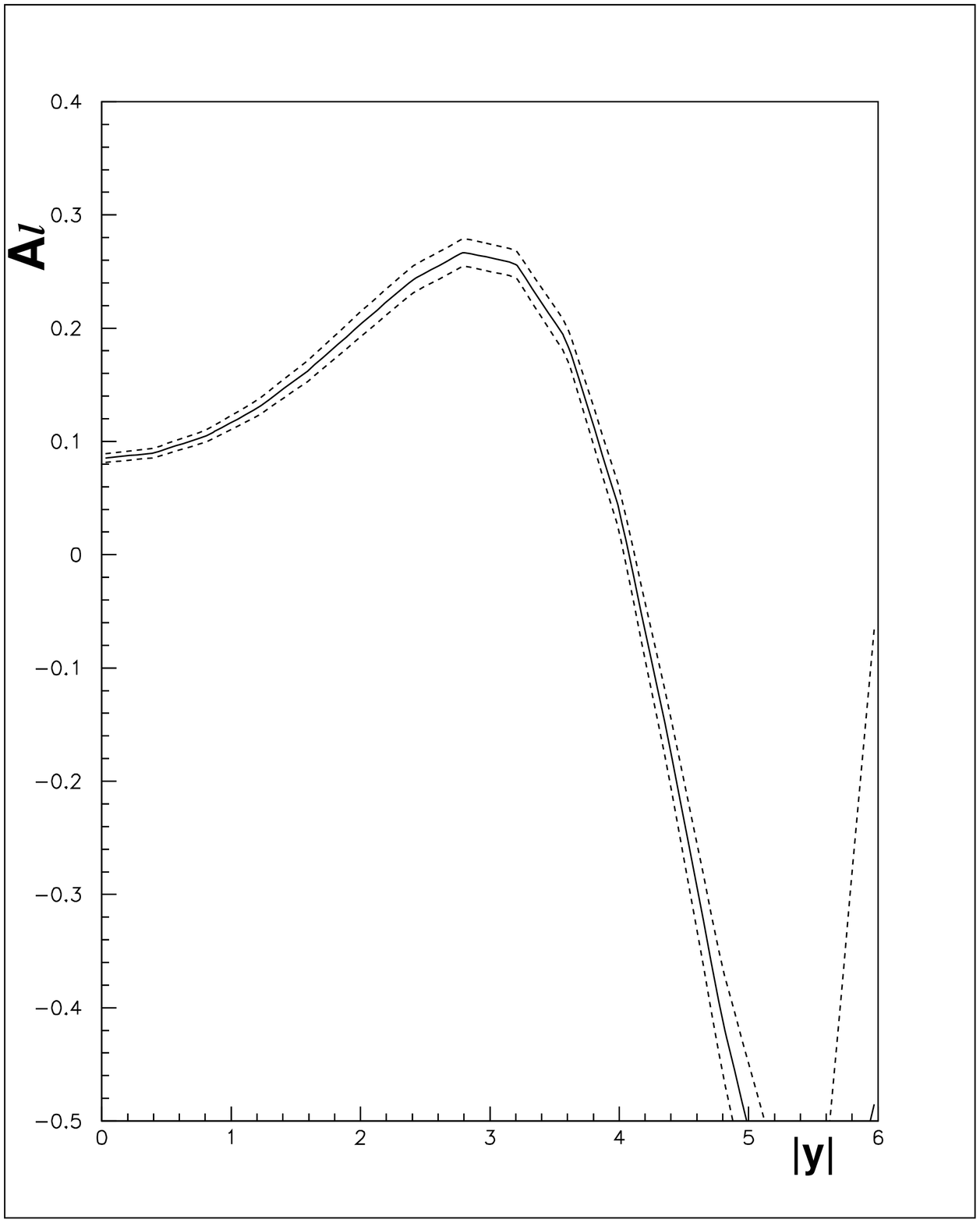,width=0.3\textwidth,height=5cm}
}
\caption {Predictions for $W,Z$ production at the LHC from the CTEQ6.1 PDFs. Top row: left plot, 
the $W$ asymmetry, $A_W$; 
right plot, the ratio, $A_{ZW}$: Bottom row: left plot, 
decay $e^+$ rapidity spectrum; middle plot, decay $e^-$ rapidity spectrum; right plot, 
lepton asymmetry, $A_e$}
\label{fig:awzwlepton}
\end{figure}

Since the PDF uncertainty feeding into the $W^+, W^-$ and $Z$ production is mostly coming from the
gluon PDF, for all three processes, there is a strong correlation in their uncertainties, which can be 
removed by taking ratios. Fig.~\ref{fig:awzwlepton} shows the $W$ asymmetry 
\[A_W = (W^+ - W^-)/(W^+ + W^-).\] for CTEQ6.1 PDFs, which have the largest uncertainties of published 
PDF sets. The PDF uncertainties on the asymmetry are very small in the measurable rapidity range. 
An eigenvector decomposition indicates that 
sensitivity to high-$x$ $u$ and $d$ quark flavour distributions is now evident at large $y$. 
Even this residual flavour 
sensitivity can be removed by taking the ratio \[A_{ZW} = Z/(W^+ +W^-)\] as also shown in 
Fig.~\ref{fig:awzwlepton}. 
This quantity is almost independent of PDF uncertainties. These quantities have been suggested as 
benchmarks for our understanding of Standard Model Physics at the LHC. However, 
whereas the $Z$ rapidity distribution can be fully reconstructed from its decay leptons, 
this is not possible for the $W$ rapidity distribution, because the leptonic decay channels 
which we use to identify the $W$'s have missing neutrinos. Thus we actually measure the $W$'s 
decay lepton rapidity spectra rather than the $W$ rapidity spectra. The lower half of  Fig.~\ref{fig:awzwlepton} 
shows the rapidity spectra for positive and 
negative leptons from $W^+$ and $W^-$ decay and the lepton asymmetry, \[A_l = (l^+ - l^-)/(l^+ + l^-).\] 
A cut of, $p_{tl} > 25$~GeV, has been applied on the decay lepton, since it will not be possible to
trigger on 
leptons with small $p_{tl}$. A particular lepton rapidity can be fed from a range 
of $W$ rapidities so that the contributions of partons at different $x$ values is smeared out 
in the lepton spectra, but the broad features of the $W$ spectra and the sensitivity to the gluon parameters 
remain. The lepton asymmetry shows the change of sign at large $y$ which is characteristic of the $V-A$ 
structure of the lepton decay. The cancellation of the 
uncertainties due to the gluon PDF is not so 
perfect in the lepton asymmetry as in the $W$ asymmetry. Nevertheless in the 
measurable rapidity range sensitivity to PDF parameters is small. Correspondingly, the PDF uncertainties are 
also small ($~4\%$) and this quantity provides a suitable Standard Model benchmark. 

In summary, these preliminary investigations indicate that PDF uncertainties on 
predictions for the $W,Z$ rapidity 
spectra, using standard PDF sets which describe all modern data, have 
reached a precision of $\sim 8\%$. This may be good enough to consider 
using these processes as luminosity monitors. The predicted precision on ratios such as the 
lepton ratio, $A_l$,  is better ($\sim 4\%$) and this measurement may be used as a SM benchmark. 
It is likely that this current level of uncertainty will have improved before the LHC turns on-
see the contribution of C. Gwenlan to these proceedings. 
The remainder of this contribution will be concerned with the question: how accurately can we  
measure these quantities and can we use the early LHC data 
to improve on the current level of uncertainty.

\section{k-factor and PDF re-weighting} 
To investigate how well we can really measure $W$ production we need to generate samples 
of Monte-Carlo (MC) data and pass them through a simulation of a detector. 
Various technical problems 
arise. Firstly, many physics studies are done with the HERWIG (6.505)\cite{herwig}, 
which generates events at LO with parton showers to account for higher order effects. 
Distributions can be corrected from LO to NLO by k-factors which are applied as a function of the 
variable of interest. The use of HERWIG is gradually being superceded by MC@NLO (2.3)\cite{mcnlo} 
but this is 
not yet implemented for all physics processes. Thus it is necessary to 
investigate how much bias is introduced by using HERWIG with k-factors. Secondly, to simulate the 
spread of current PDF uncertainties, it is necessary to run the MC with all of the eigenvector 
error sets of the PDF of interest. This would be unreasonably time-consuming. Thus the technique 
of PDF reweighting has been investigated.

\begin{figure}[tbp] 
\vspace{-1.0cm}
\centerline{
\epsfig{figure=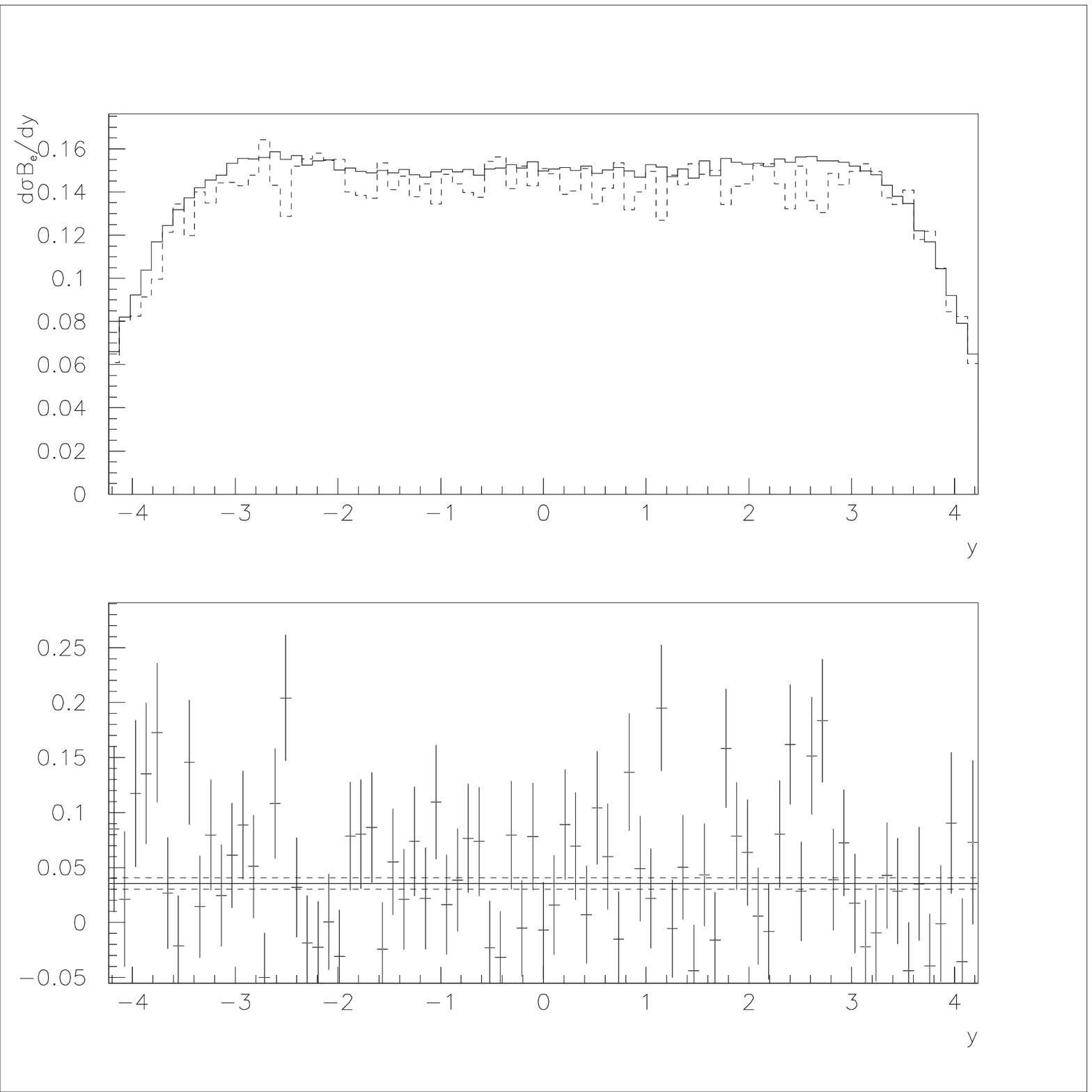,width=0.3\textwidth}
\epsfig{figure=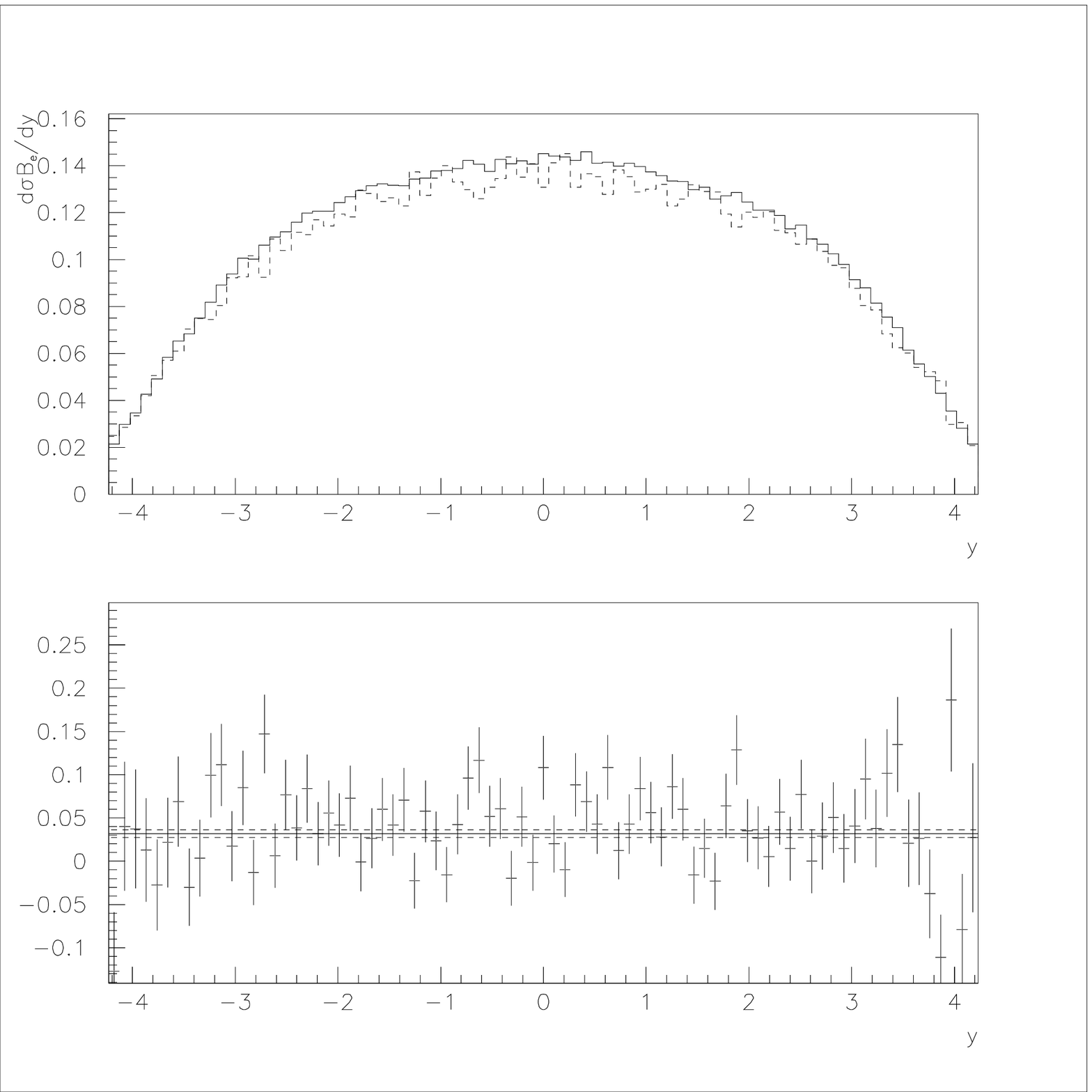,width=0.3\textwidth}
\epsfig{figure=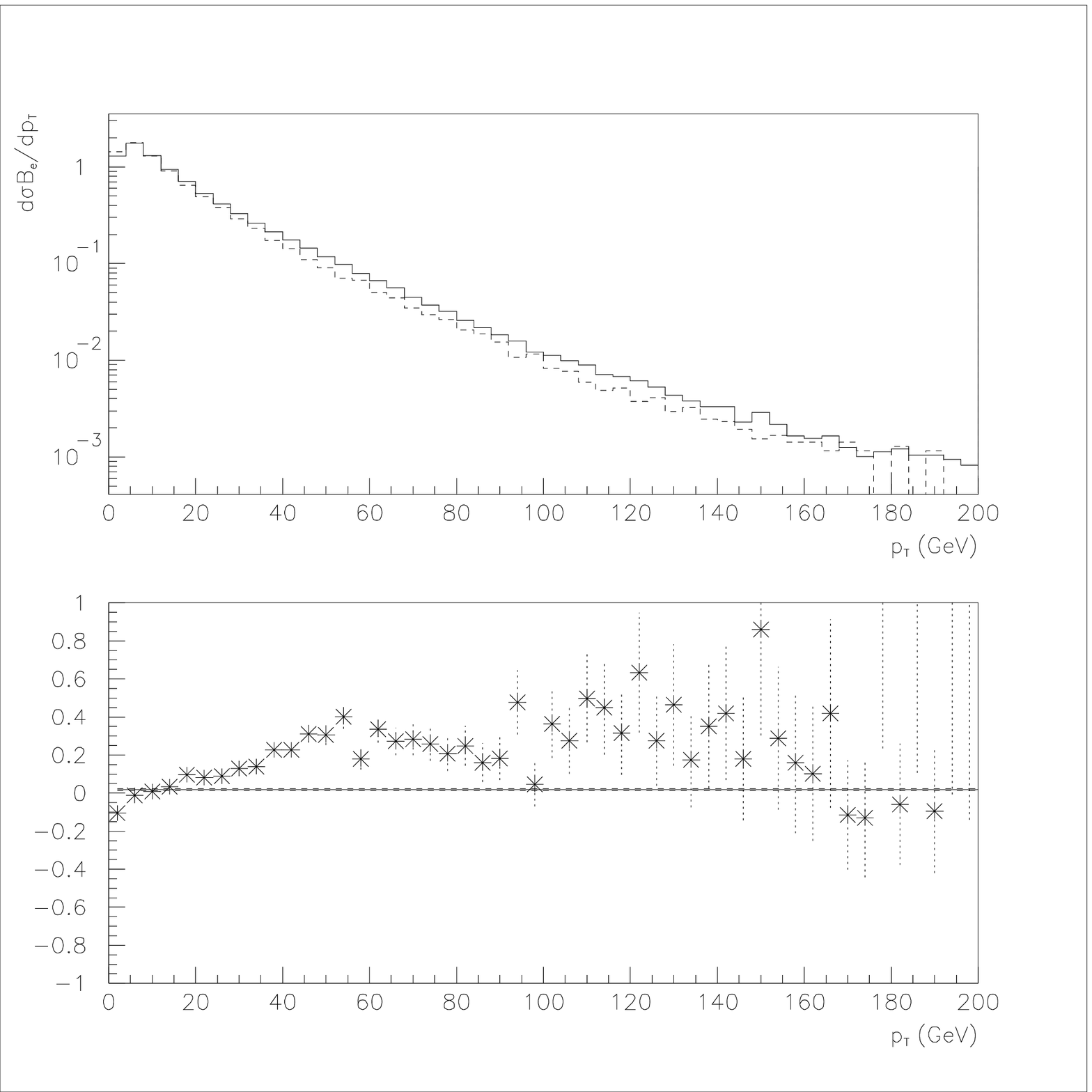,width=0.3\textwidth}
}
\caption {Top Row: $W$ rapidity and $p_t$ spectra for events generated with HERWIG + k-Factors (full line), 
compared to those generated by MC@NLO (dashed line); left plot $W^+$ rapidity; middle plot $W^-$ rapidity; 
right plot $W^-$ $p_t$. Bottom row:
the fractional differences of the spectra generated by HERWIG + k-factors and those generated by MC@NLO. 
The full line represents the weighted mean of these difference spectra and the dashed lines show 
its uncertainty }
\label{fig:kfactor}
\end{figure}

One million $W \to e \nu_e$ events were generated using HERWIG (6.505). This corresponds to 43 hours of LHC running at low 
luminosity, $10 fb^{-1}$. These events are split into $W^+$ and $W^-$ events according to 
their Standard Model 
cross-section rates, $58\%$: $42\%$ (the exact split depends on the input PDFs). These events are then 
weighted with k-factors, which are analytically calculated as the ratio of the NLO to LO 
cross-section as a function of rapidity for the same input PDF~\cite{Stirling}. 
The resultant rapidity 
spectra for $W^+, W^-$ are compared to rapidity spectra for $\sim 107,700$ events 
generated using MC@NLO(2.3) in Fig~\ref{fig:kfactor}\footnote{In MC@NLO the hard emissions are treated 
by NLO 
computations, whereas soft/collinear emissions are handled by the MC simulation. In the matching procedure a 
fraction of events with negative weights is generated to avoid double counting. The event  weights must be 
applied to the generated number of events 
before the effective number of events can be converted to an equivalent luminosity. The figure given is the 
effective number of events.}. The MRST02 PDFs were used for this investigation. 
The accuracy of this study is limited by the statistics 
of the MC@NLO generation. Nevertheless it is clear that HERWIG with k-factors does a good job of mimicking 
the NLO rapidity spectra. However, the normalisation is too high by $3.5\%$. This is not suprising since, 
unlike the analytic code, HERWIG is not a purely LO calculation, parton showering is also included. 
This normalisation difference is not too crucial since
in an analysis on real data the MC will only be used to correct data from the detector level to the 
generator level. For this purpose, 
it is essential to model the shape of spectra to understand the effect of experimental cuts and smearing
but not essential to model the overall normalisation perfectly. However, one should note that 
HERWIG with k-factors is not so successful in modelling the shape of the $p_t$ spectra, as shown in the 
right hand plot of
Fig.~\ref{fig:kfactor}. This is hardly surprising, since at LO the $W$ have no $p_t$ and non-zero $p_t$ 
for HERWIG is generated by parton showering, whereas for MC@NLO non-zero $p_t$ originates from additional higher order processes which cannot be scaled from LO, where they are not present. 

\begin{figure}[tbp] 
\vspace{-1.0cm}
\centerline{
\epsfig{figure=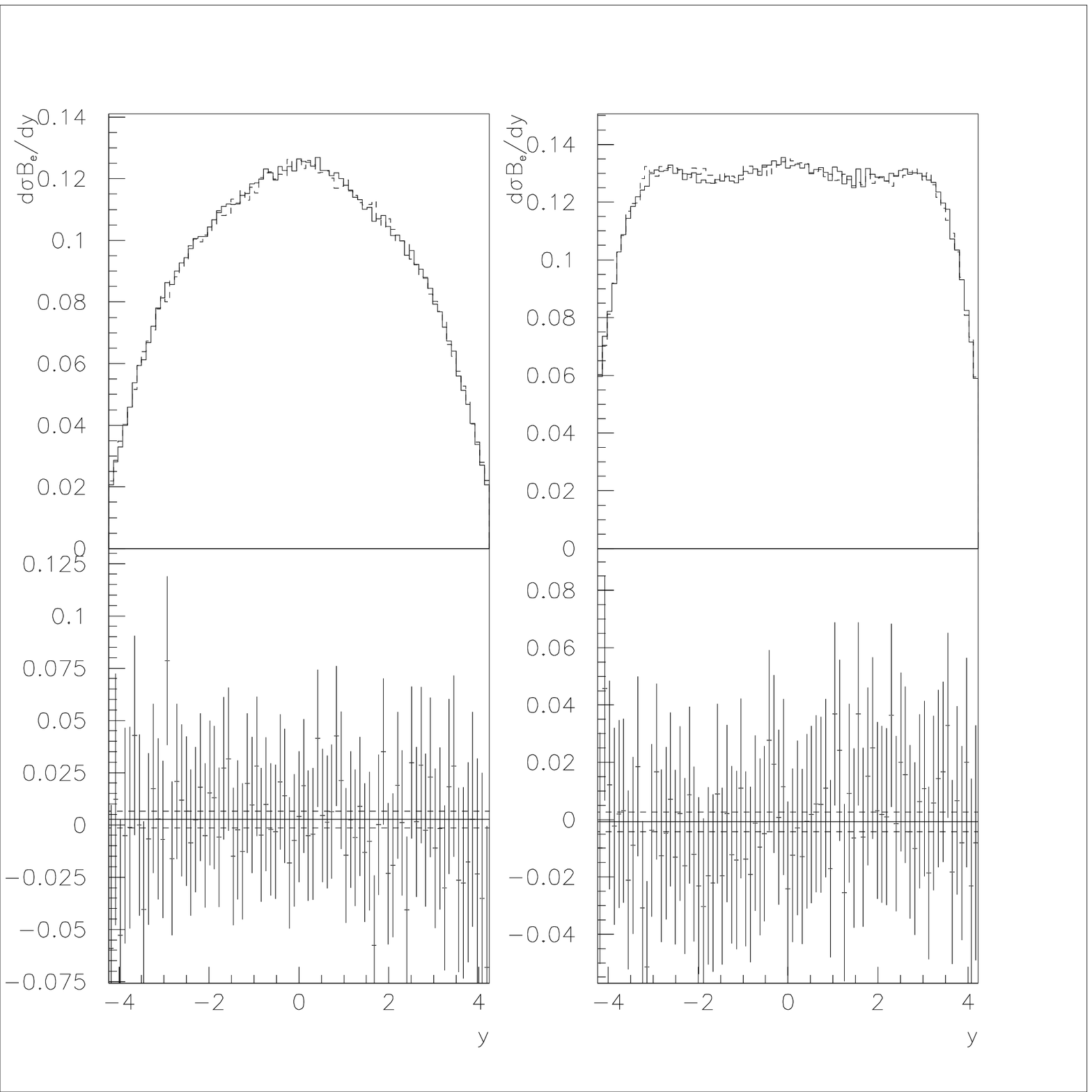,height=7cm}
\epsfig{figure=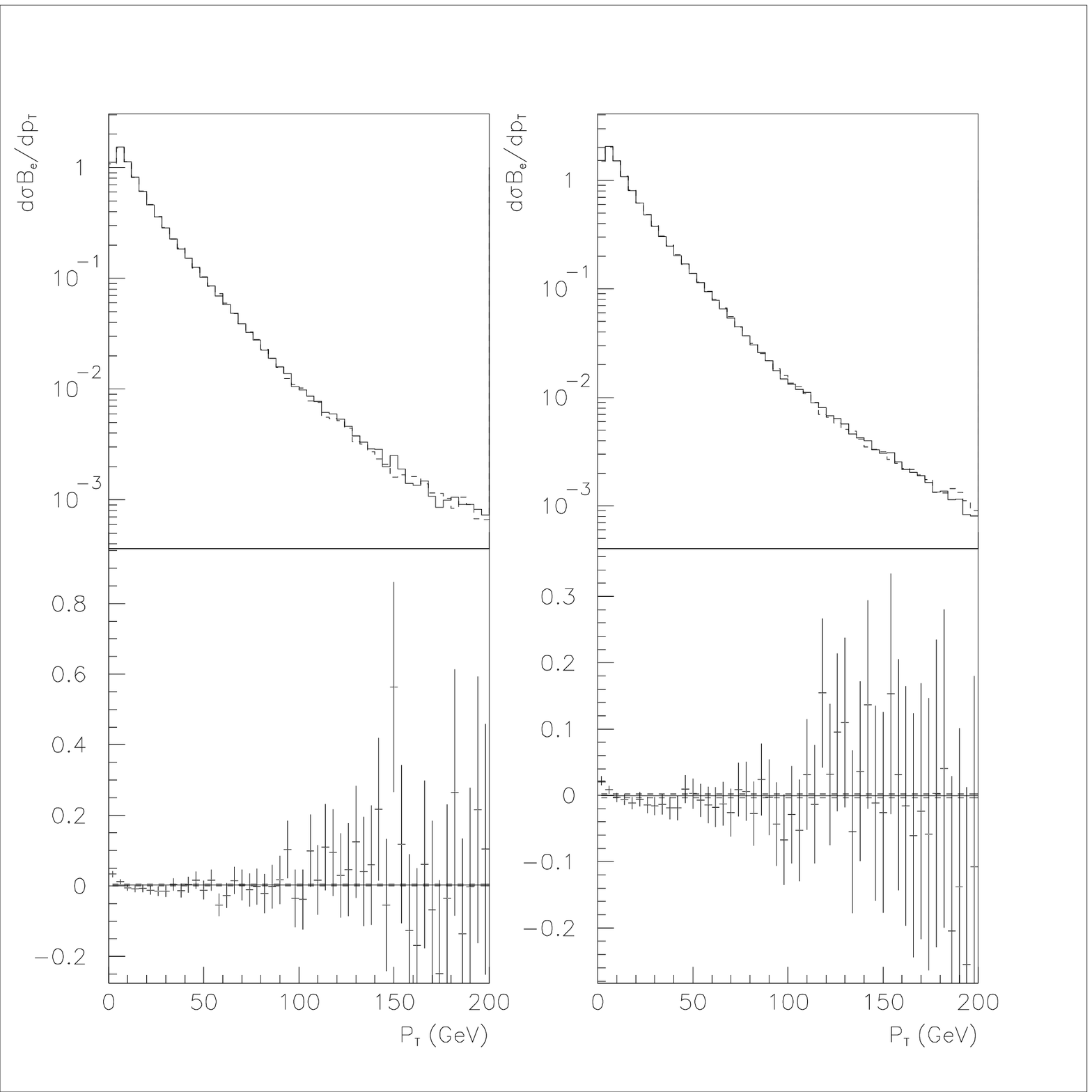,height=7cm}
}
\caption {Left side: 
$W^-$ (left) and $W^+$ (right) rapidity spectra, for events generated with MRST02 PDFs reweighted to 
CTEQ6.1 PDFs (full line), compared to events generated directly with CTEQ6.1 PDFs (dashed line). 
The fractional difference between these spectra are also shown
beneath the plots. The full line represents the weighted mean of these difference spectra and the dashed 
lines show its uncertainty. Right side: the same for $p_t$ spectra.}
\label{fig:pdfreweight}
\end{figure}

Suppose we generate $W$ events with a particular PDF set:
PDF set 1. Any one event has the hard scale, $Q^2 = M_W$, and two primary partons of 
flavours $flav_1$ and $flav_2$, 
with momentum fractions $x_1, x_2$ according to the distributions of PDF set 1. These momentum 
fractions are applicable to the hard process before the parton showers are implemented in backward
evolution in the MC. One can then evaluate the probability of picking up the same flavoured 
partons with the same momentum fractions from an alternative PDF set, PDF set 2, 
at the same hard scale. Then the event weight is given by
\begin{equation}
  \rm{PDF re-weight} = \frac{f_{PDF_2}(x_1,flav_1, Q^2). f_{PDF_2} (x_2, flav_2, Q^2)}
{ f_{PDF_1}(x_1,flav_1, Q^2). f_{PDF_1} (x_2, flav_2, Q^2)}
\end{equation}
where $xf_{PDF}(x,flav,Q^2)$ is the parton momentum distribution for flavour, $flav$, 
at scale, $Q^2$, and momentum fraction, $x$.
Fig.~\ref{fig:pdfreweight} compares the $W^+$ and $W^-$ spectra for a million events 
generated using MRST02 as PDF set 1 and re-weighting  to CTEQ6.1 as PDF set 2, 
with a million events which are directly generated with CTEQ6.1.  Beneath the spectra the fractional difference
between these distributions is shown. These difference spectra show that the reweighting is good 
to better than $1\%$, and there is no evidence of a $y$ dependent bias. 
This has been checked for reweighting between MRST02, CTEQ6.1 and ZEUS-S 
PDFs. Since the uncertainties of any one analysis are similar in size to the differences between 
the analyses it is clear that the technique can be used to produce spectra for the eigenvector 
error PDF sets of each analysis and thus to simulate the full PDF uncertainties from a single 
set of MC generated events.  
Fig.~\ref{fig:pdfreweight} also shows a similar comparison for $p_t$ spectra. 

\section{Background Studies}
\label{sec:bgd}
To investigate the accuracy with which $W$ events can be measured at the LHC
it is necessary to make an estimate of the importance of 
background processes. We focus on $W$ events which are identified through their decay to the 
$W \rightarrow e~\nu_e$ channel. There are several processes which can be misidentified as
$W \rightarrow e \nu_e$. These are: $W \rightarrow \tau \nu_\tau$, with $\tau$ decaying to the electron 
channel; $Z \rightarrow \tau^+ \tau^-$ with at least one $\tau$ decaying to the electron channel 
(including the case when both $\tau$'s decay to the electron channel, but one 
electron is not identified); $Z \rightarrow e^+ e^-$ with one electron not identified. We have generated one 
million events for each of these background processes, 
using HERWIG and CTEQ5L, and compared them to one million signal events generated with CTEQ6.1. 
We apply event selection criteria designed to eliminate the background preferentially. These criteria are:
\begin{itemize}
\item ATLFAST cuts (see Sec.~\ref{sec:gendet})
\item pseudorapidity, $|\eta| <2.4$, to avoid bias at the edge of the measurable rapidity range
\item  $p_{te} > 25$ GeV, high $p_t$ is necessary for electron triggering 
\item  missing $E_t > 25$ GeV, the $\nu_e$ in a signal event will have a correspondingly large missing $E_t$
\item  no reconstructed jets in the event with $p_t > 30$ GeV, to discriminate against QCD background 
\item  recoil on the transverse plane $p_t^{recoil} < 20$ GeV, to discriminate against QCD background
\end{itemize}
Table~\ref{tab:bgd} gives the percentage of background with respect to signal, calculated using the known 
relative cross-sections of these processes, as each of these cuts is 
applied. 
After, the cuts have been applied the background from these processes is negligible. However,
there are limitations on this study from the fact that in real data there will be further QCD backgrounds 
from $2 \rightarrow 2$ processes involving $q,\bar{q},g$ in which a final state $\pi_0 \rightarrow \gamma
\gamma$ decay mimics a single electron. A preliminary study applying the selection criteria to MC 
generated QCD events suggests 
that this background is negligible, but the exact level of QCD background cannot be accurately estimated 
without passing a very large 
number of events though a full detector simulation, which is beyond the scope of the current contribution.
\begin{table}[tbp]
\vspace{-1.0cm}
\caption{Reduction of signal and background due to cuts}
\centerline{\small
\begin{tabular}{c|cc|cc|cc|cc}\\
 \hline
Cut & ~~~$W \rightarrow e \nu_e$ &  & 
~~~$Z \rightarrow \tau^+ \tau^-$ &  &
~~~$Z \rightarrow e^+ e^- $ &  &
~~~$W \rightarrow \tau \nu_\tau$ &  
\\
 & $e^+$ & $e^-$ &$e^+$ & $e^-$ &$e^+$ & $e^-$ &$e^+$ & $e^-$ 
\\ 
 \hline
 ATLFAST cuts& 382,902 & 264,415  & $5.5\%$ & $7.9\%$ & $34.7\%$ & $50.3\%$ & $14.8\%$ & $14.9\%$\\
 $|\eta| < 2.4$ & 367,815 & 255,514  & $5.5\%$ & $7.8\%$ & $34.3\%$ & $49.4\%$ & $14.7\%$ & $14.8\%$\\
 $p_{te} > 25$ GeV& 252,410 & 194,562  & $0.6\%$ & $0.7\%$ & $12.7\%$ & $16.2\%$ & $2.2\%$ & $2.3\%$\\
 $p_{tmiss} > 25$ GeV& 212,967 & 166,793  & $0.2\%$ & $0.2\%$ & $0.1\%$ & $0.2\%$ & $1.6\%$ & $1.6\%$\\
 No jets with $P_t > 30$ GeV& 187,634 & 147,415  & $0.1\%$ & $0.1\%$ & $0.1\%$ & $0.1\%$ & $1.2\%$ & $1.2\%$\\
$p_t^{recoil} < 20$ GeV& 159,873 & 125,003  & $0.1\%$ & $0.1\%$ & $0.0\%$ & $0.0\%$ & $1.2\%$ & $1.2\%$\\
 \hline\\
\end{tabular}}
\label{tab:bgd}
\end{table}

\section{Charge misidentification}

\begin{figure}[tbp] 
\vspace{-0.8cm}
\centerline{
\epsfig{figure=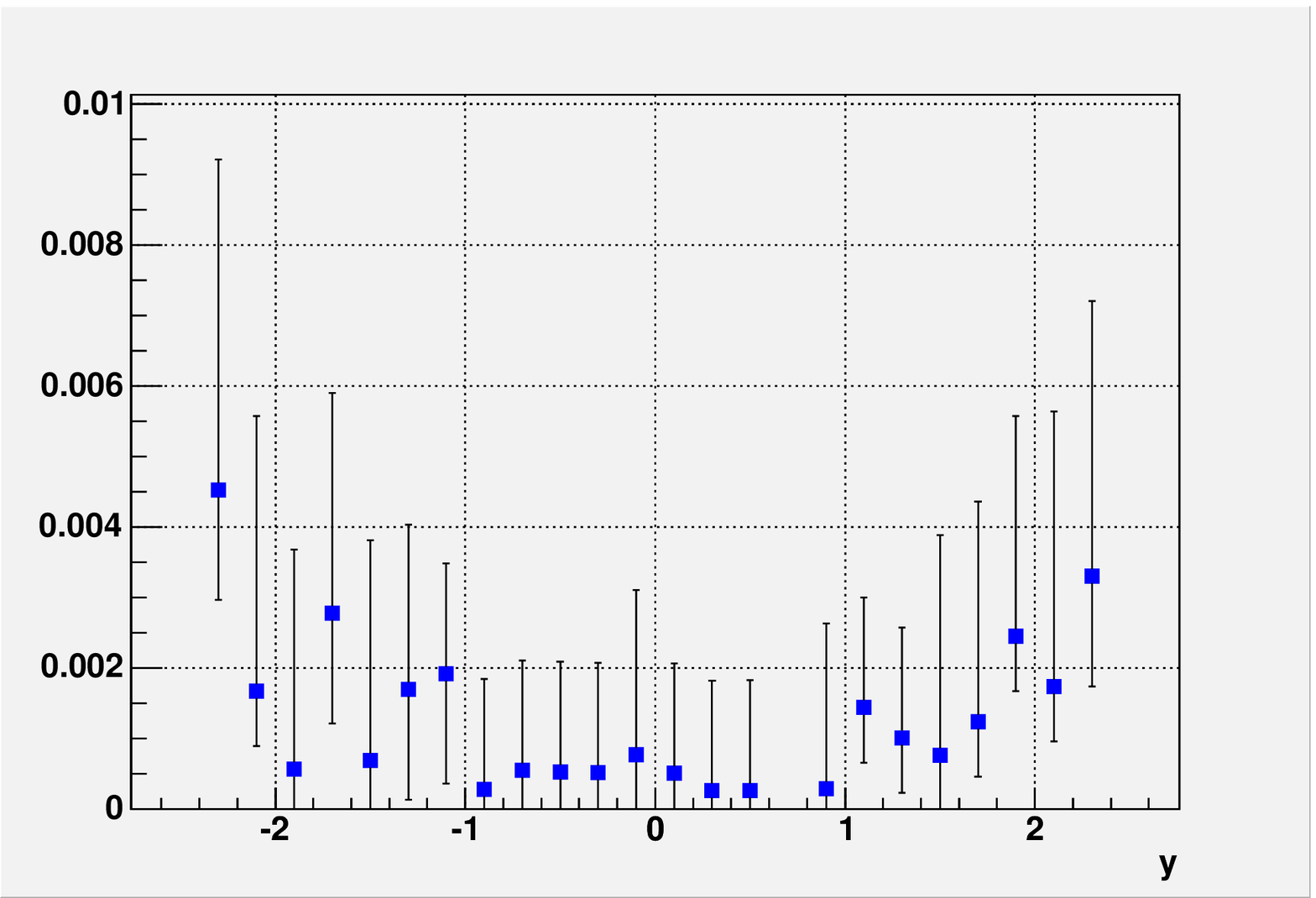,width=0.4\textwidth}
\epsfig{figure=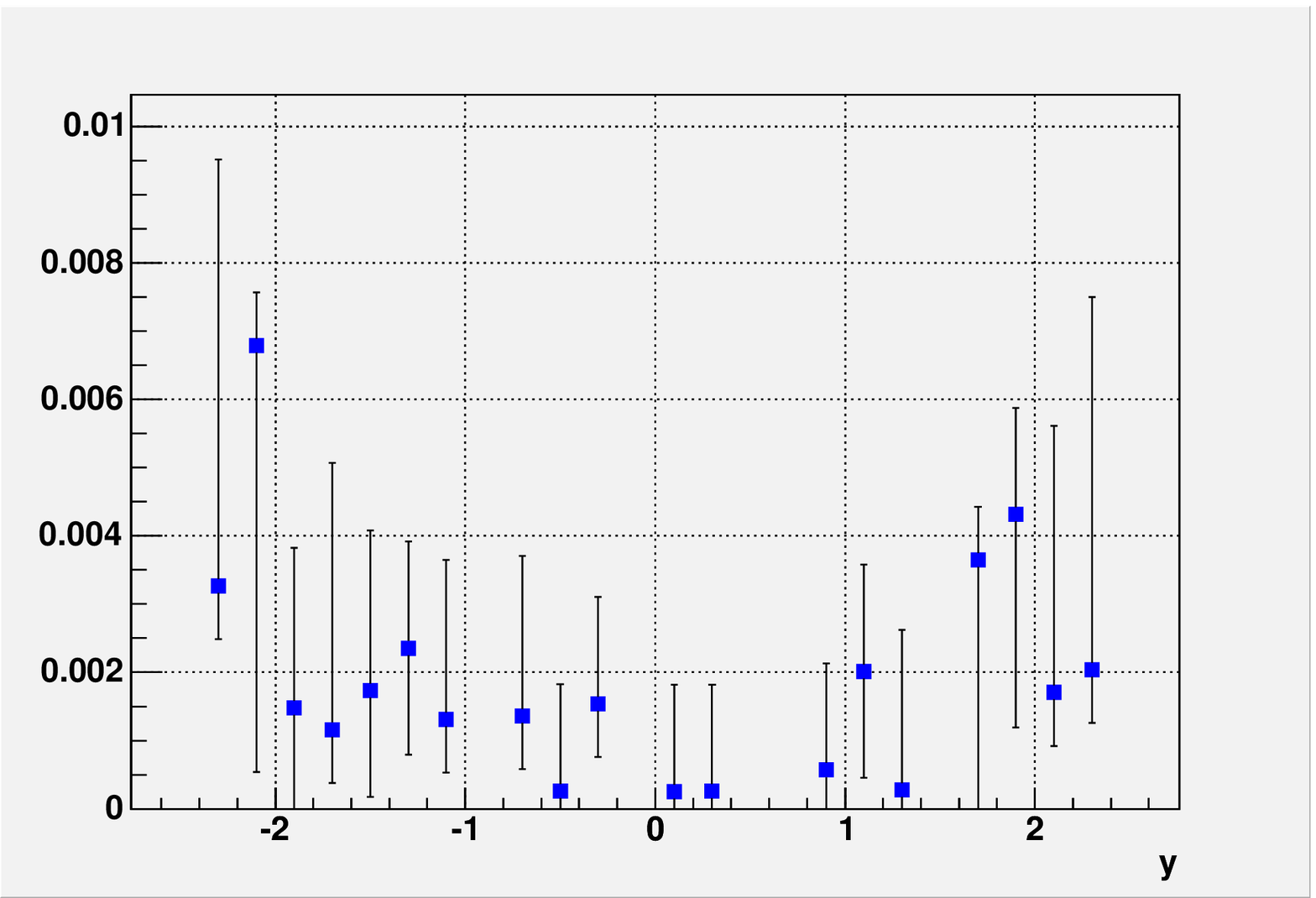,width=0.4\textwidth}
}
\caption {The rates of charge misidenttification as a function of rapidity for $e^-$ misidentified 
as $e^+$ (left), $e^+$ misidentifed as $e^-$ (right).
}
\label{fig:misid}
\end{figure}

Clearly charge misidentification could distort the lepton rapidity spectra and dilute the asymmetry $A_l$.
\[
 A_{true} = \frac{ A_{raw} - F^+ + F^-}{1 -F^- - F^+}
\]
where $A_{raw}$ is the measured asymmetry, $A_{true}$ is the true asymmetry, $F^-$ is the rate of true $e^-$
misidentified as $e^+$ and $F^+$ is the rate of true $e^+$ misidentified as $e^-$. To make an estimate of 
the importance of charge misidentification we use a sample of $Z \rightarrow e^+ e^-$ events generated by 
HERWIG with CTEQ5L and passed through a full simulation of the ATLAS detector. Events with two
or more charged electromagnetic objects in the EM calorimeter are then selected and subject to the cuts;
$|\eta| < 2.5$, $p_{te} > 25$ GeV, as 
usual and, $E/p < 2$, for bremsstrahlung rejection. We then look for the charged electromagnetic pair with 
invariant mass closest to $M_Z$ and impose the cut, $60 < M_Z < 120$ GeV. 
Then we tag the charge of the better 
reconstructed lepton of the pair and check to see if the charge of the second lepton is the
 same as the first. Assuming that the pair really came from the decay of the $Z$ this gives us a measure of 
charge misidentification. Fig~\ref{fig:misid} show the misidentification rates $F^+$, $F^-$ as functions of 
pseudorapidity\footnote{These have been corrected for the small possibility that the better reconstructed lepton 
has had its charge misidentified as follows. In the central region, $|\eta| < 1$, assume the same probability 
of misidentification of the first and second leptons, in the more forward regions assume the same rate of 
first lepton misidentification as in the central region.}.
 These rates are very small. The quantity $A_l$, can be corrected 
for charge misidentification applying Barlow's method for combining asymmetric errors~\cite{Barlow}. 
The level of correction is $0.3\%$ in the central region 
and $0.5\%$ in the more forward regions. 

\section{Compare events at the generator level to events at the detector level}
\label{sec:gendet}
\begin{figure}[tbp] 
\vspace{-0.5cm}
\centerline{
\epsfig{figure=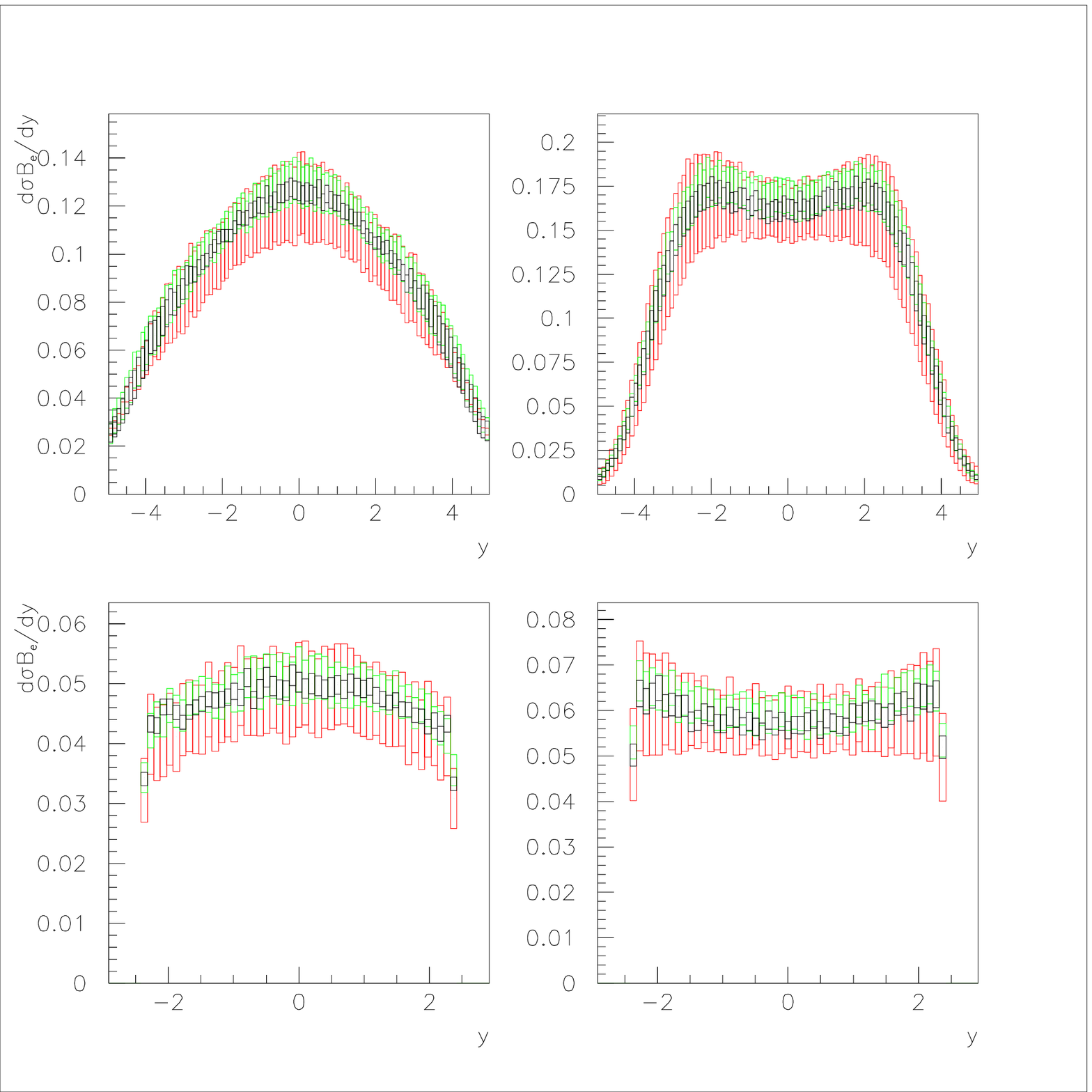,width=0.66\textwidth,height=8cm }
\epsfig{figure=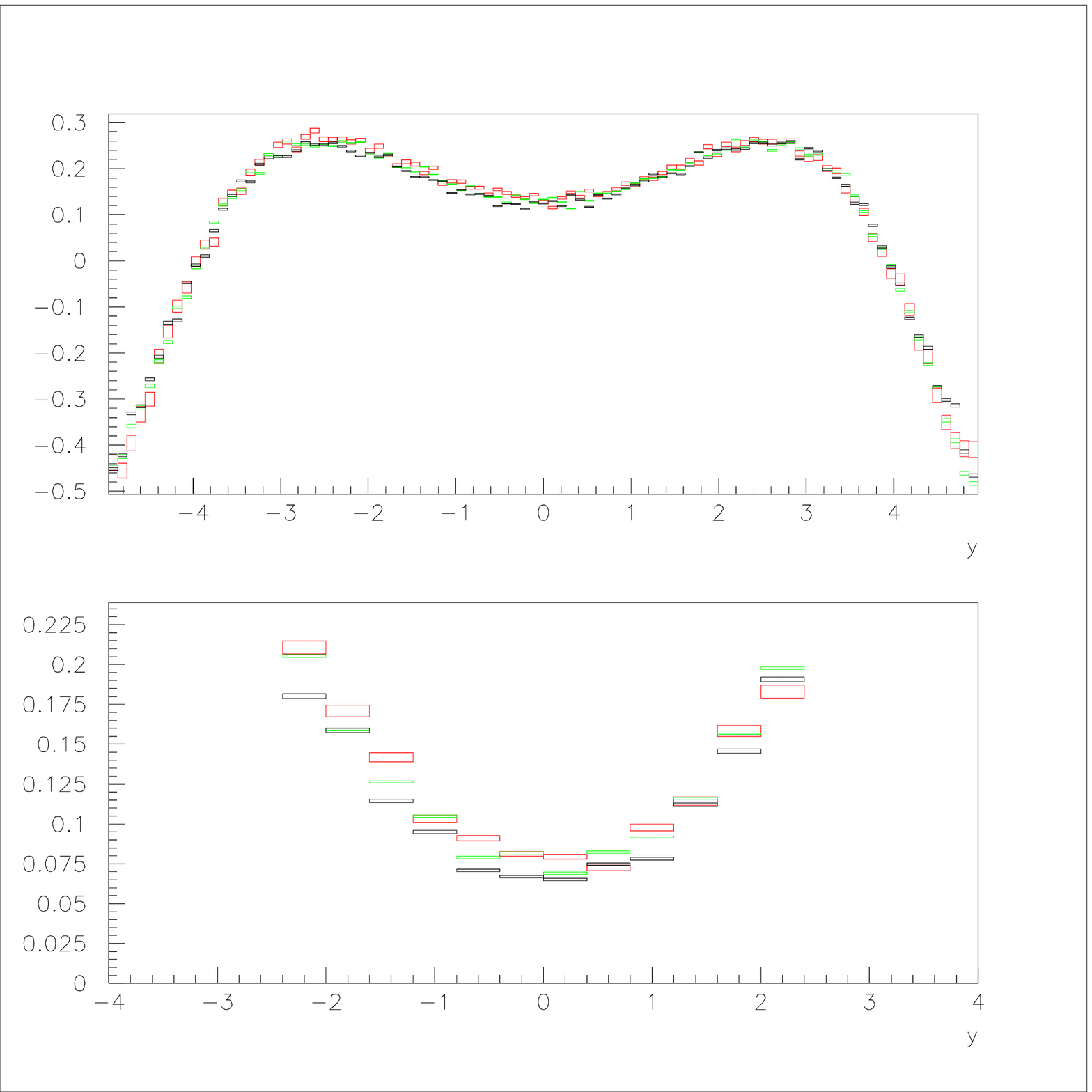,width=0.33\textwidth,height=8cm}
}
\caption {Top row: $e^-$, $e^+$ and $A_e$ rapidity spectra for the lepton from the $W$ decay, 
generated using HERWIG + k factors and CTEQ6.1 (red),
ZEUS-S (green) and MRST2001 (black) PDF sets with full uncertainties. Bottom row: the same spectra after passing 
through the ATLFAST detector simulation and selection cuts.}
\label{fig:gendet}
\end{figure}

We have simulated one million signal, $W \rightarrow e \nu_e$, events for each of the PDF sets CTEQ6.1, 
MRST2001 and ZEUS-S using HERWIG (6.505). 
For each of these PDF sets the eigenvector error PDF sets have been simulated 
by PDF reweighting and k-factors have been applied to approximate an NLO generation.
The top part of Fig.~\ref{fig:gendet}, shows the $e^{\pm}$ and $A_l$ 
spectra at this generator level, for all of the PDF sets superimposed.
The events are then passed through the ATLFAST fast simulation of the ATLAS detector. This applies
loose kinematic cuts: $|\eta| < 2.5$, $p_{te} > 5$ GeV, and electron isolation criteria. 
It also smears the 4-momenta of the 
leptons to mimic momentum dependent detector resolution. We then apply the selection cuts described in 
Sec.~\ref{sec:bgd}. The lower half of
Fig.~\ref{fig:gendet}, shows the $e^{\pm}$ and $A_l$ spectra at the detector level after application 
of these cuts, for all of the PDF sets superimposed. 
The level of precision of each PDF set, seen in the analytic calculations of Fig.~\ref{fig:mrstcteq},
 is only slightly degraded  at detector level, so that a net level of PDF 
uncertainty at central rapidity of $\sim 8\%$ is maintained. The anticipated cancellation of PDF 
uncertainties in the asymmetry spectrum is also observed, within each PDF set, and the spread between PDF sets
suggests that measurements which are accurate to better than $\sim 5\%$ could discriminate between PDF sets.

\section{Using LHC data to improve precision on PDFs}

\begin{figure}[tbp] 
\vspace{-1.5cm}
\centerline{
\epsfig{figure=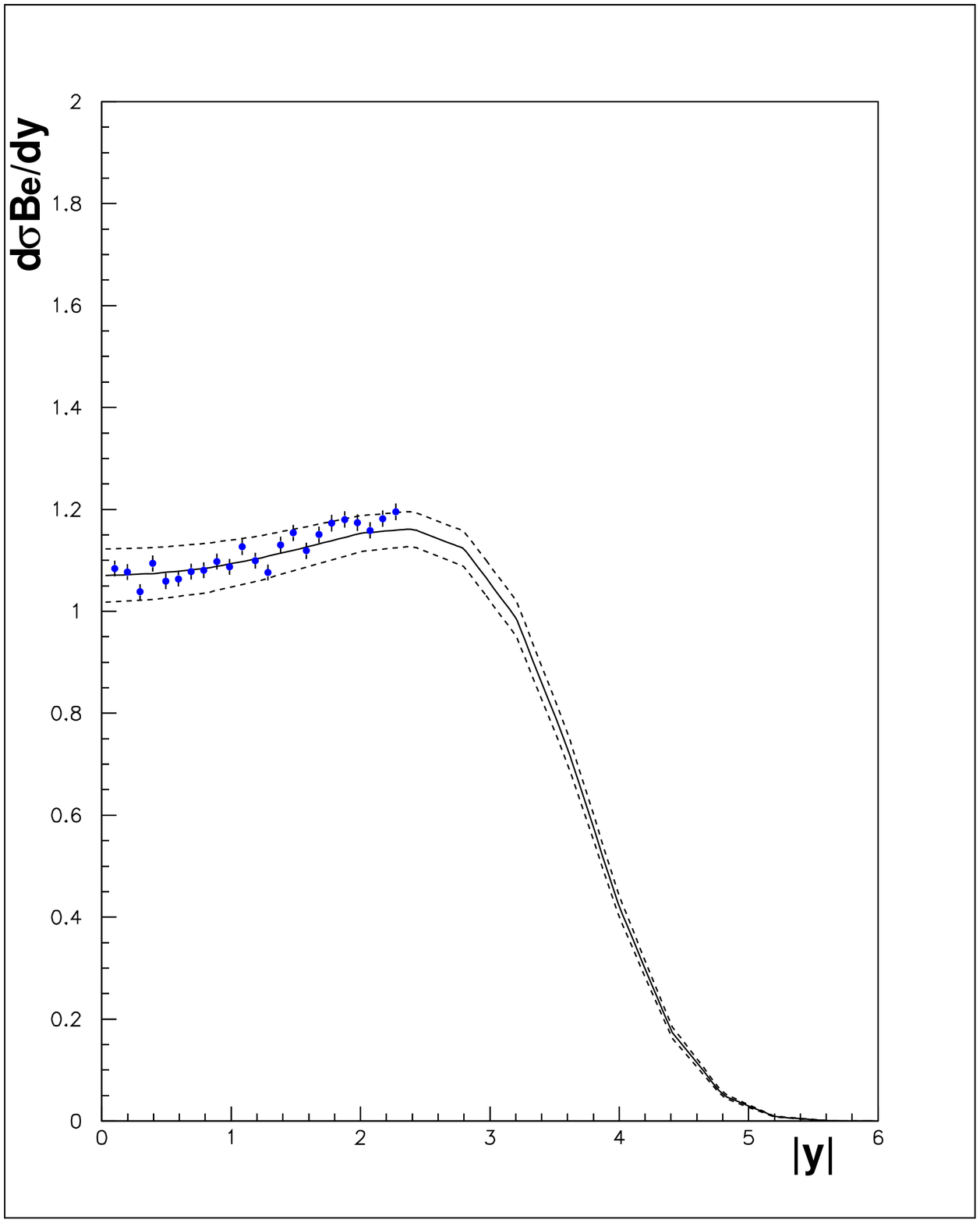,width=0.3\textwidth,height=5cm}
\epsfig{figure=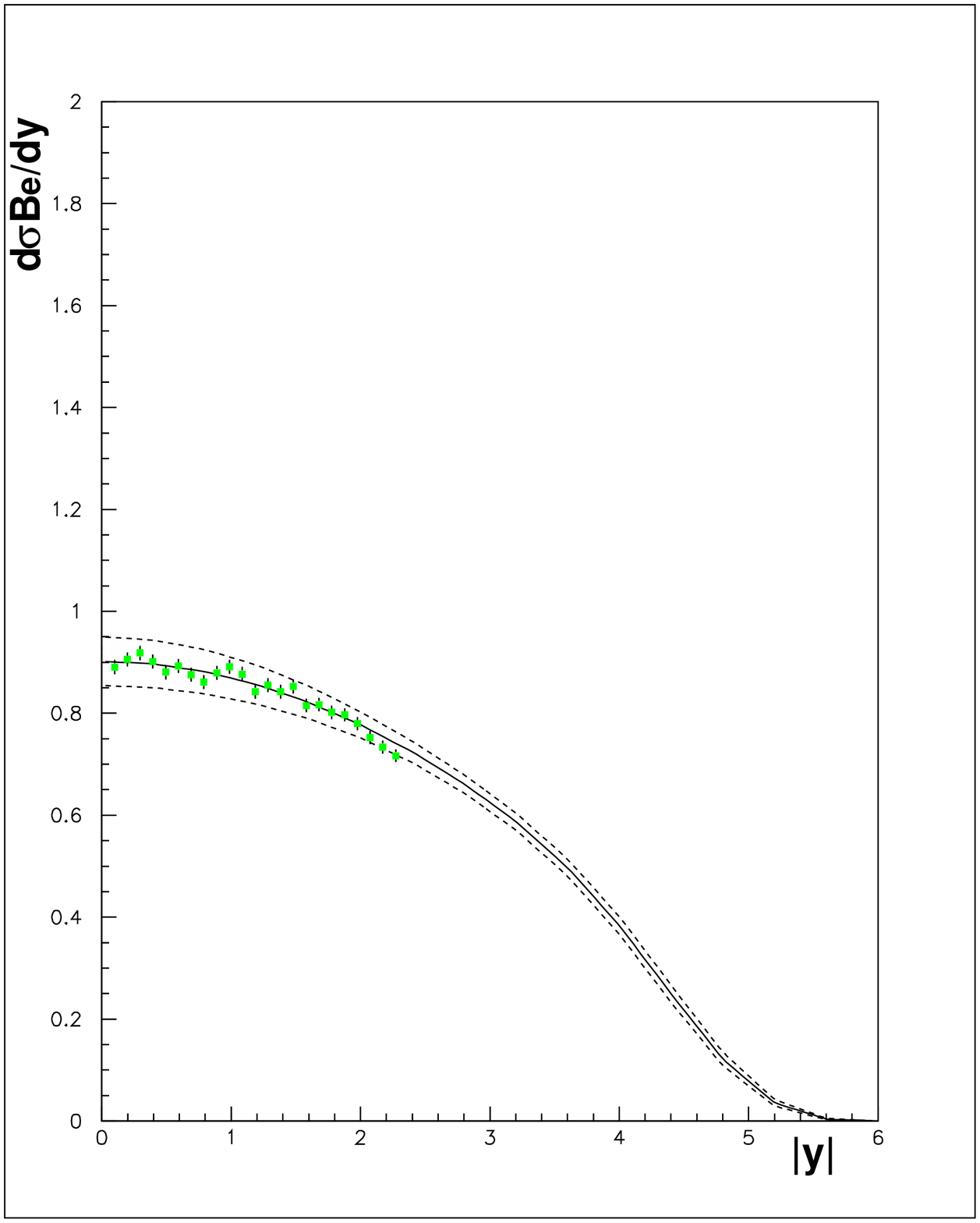,width=0.3\textwidth,height=5cm}
}
\centerline{
\epsfig{figure=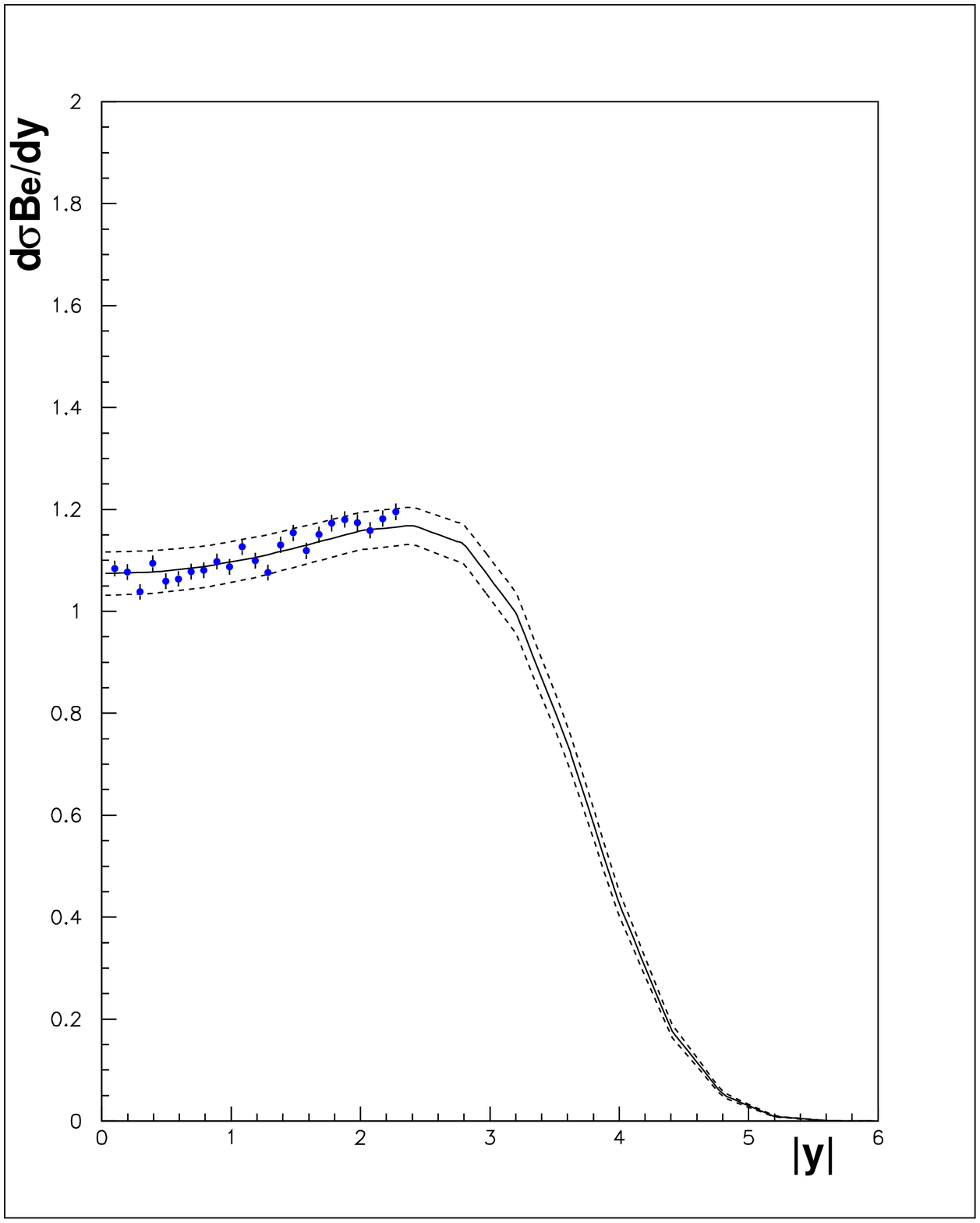,width=0.3\textwidth,height=5cm}
\epsfig{figure=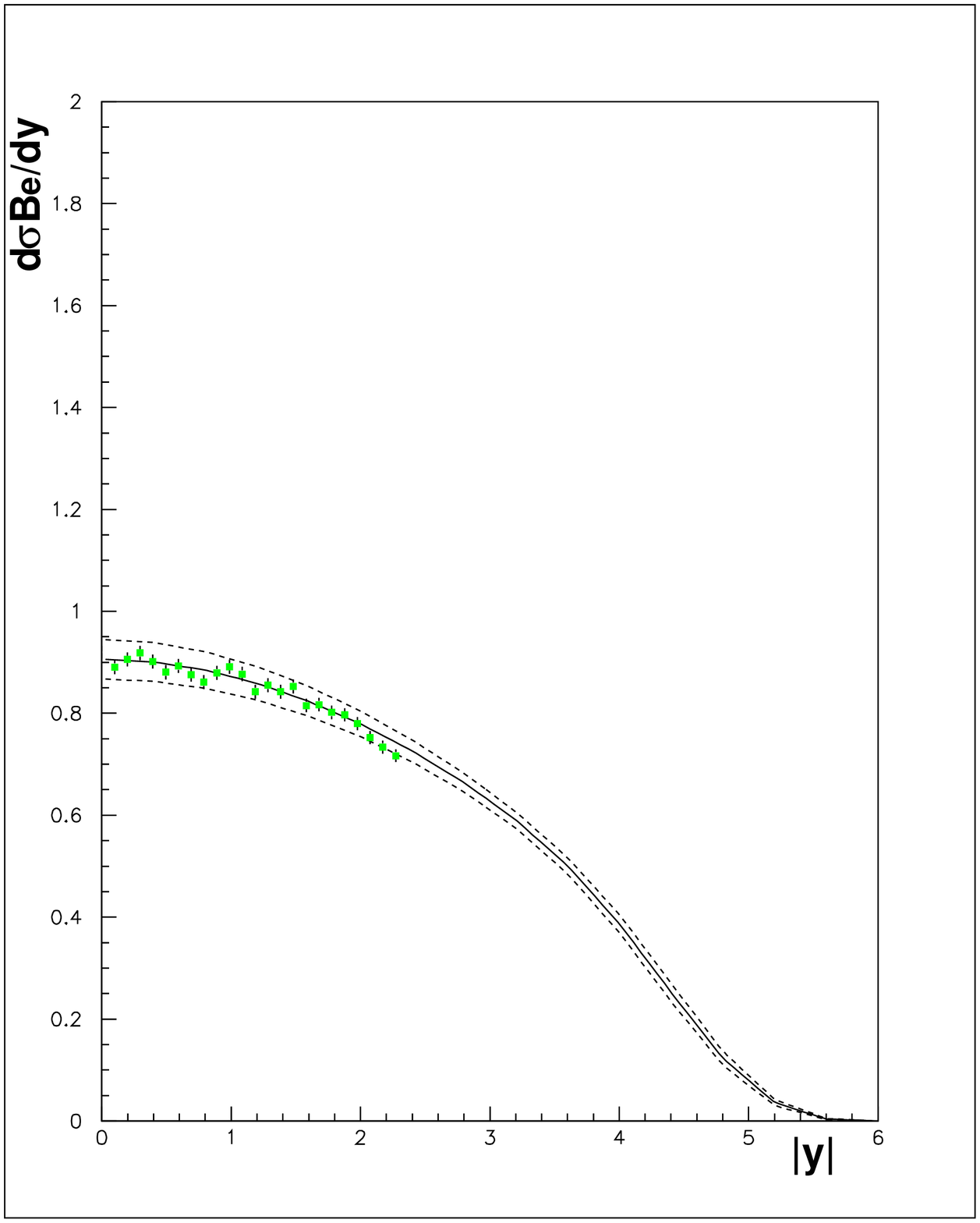,width=0.3\textwidth,height=5cm}
}
\caption {Top row: $e^+$ and $e^-$ rapidity spectra generated from ZEUS-S PDFs compared to the analytic prediction
using ZEUS-S PDFs. Bottom row: the same lepton rapidity spectra compared to the analytic 
prediction AFTER including these lepton pseudo-data in the ZEUS-S PDF fit.}
\label{fig:zeusfit}
\end{figure}

\begin{figure}[tbp] 
\vspace{-1.5cm}
\centerline{
\epsfig{figure=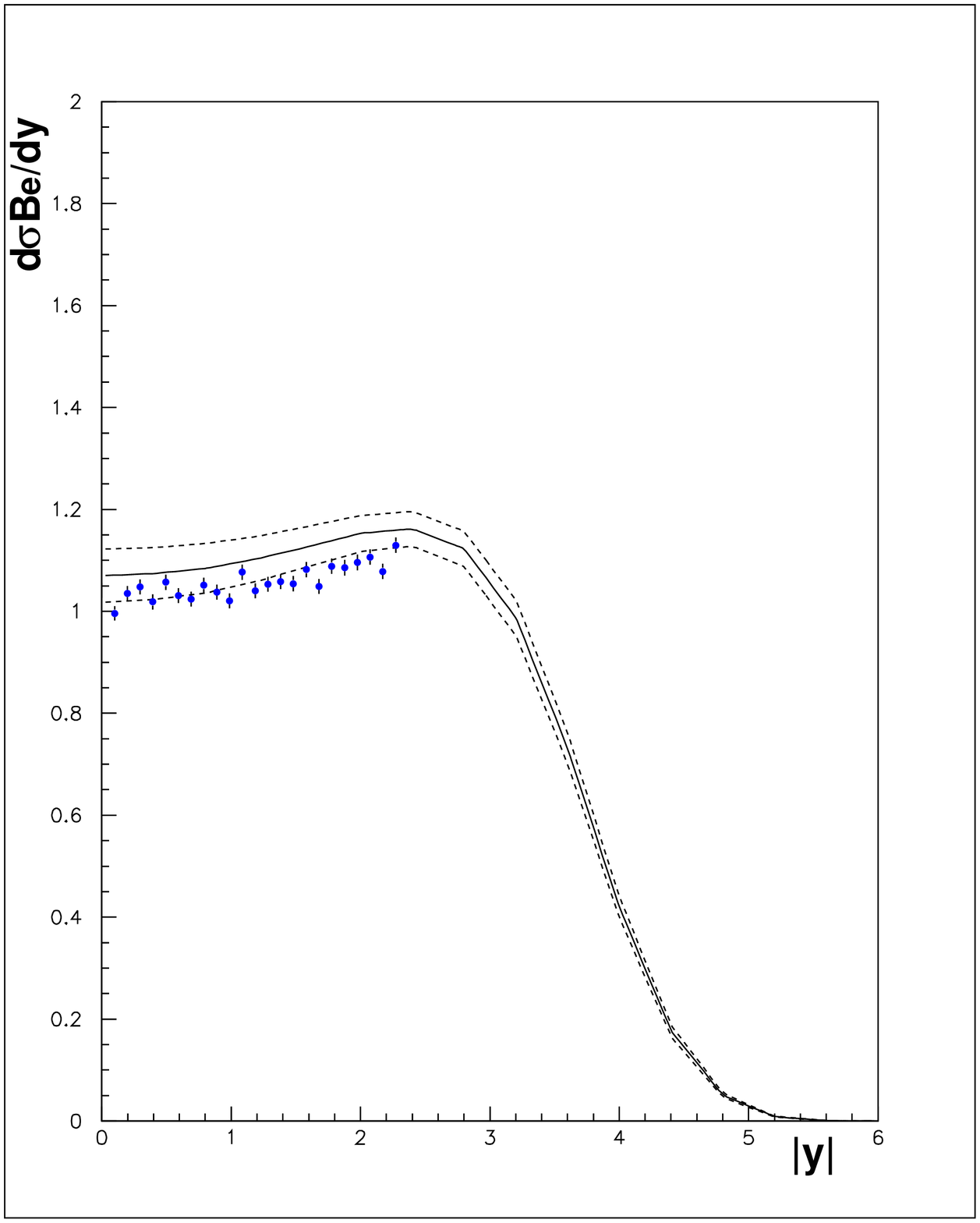,width=0.3\textwidth,height=5cm}
\epsfig{figure=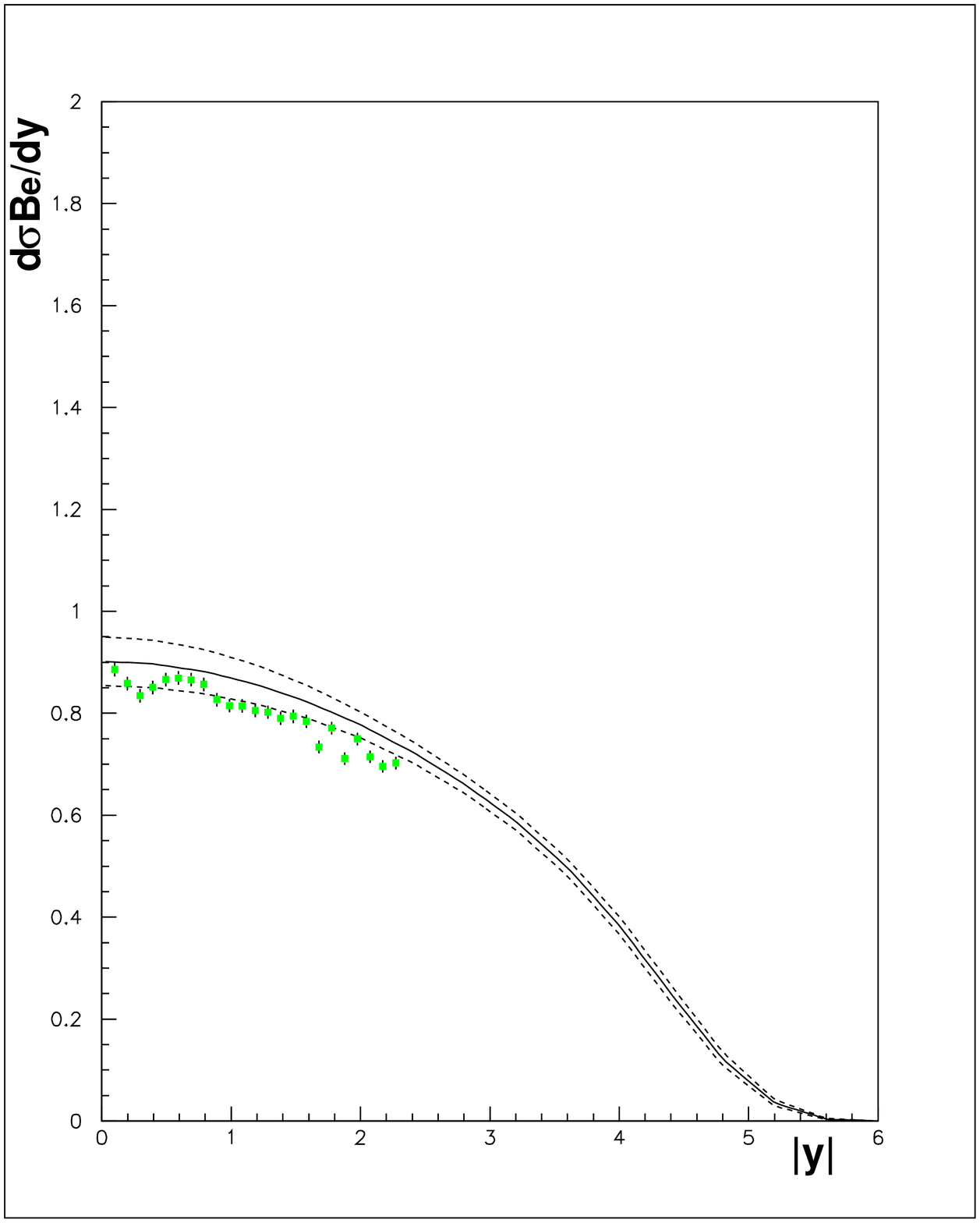,width=0.3\textwidth,height=5cm}
}
\centerline{
\epsfig{figure=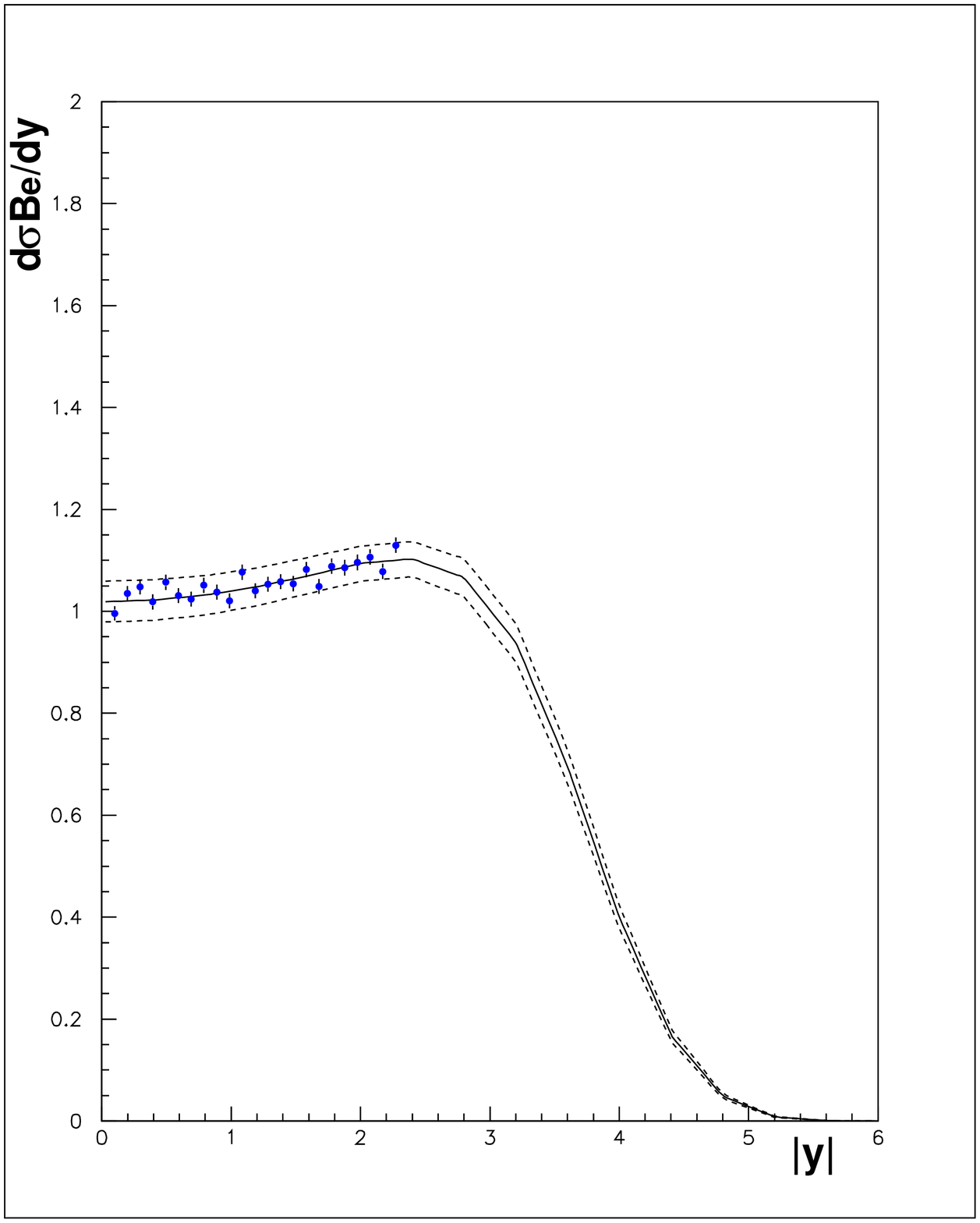,width=0.3\textwidth,height=5cm}
\epsfig{figure=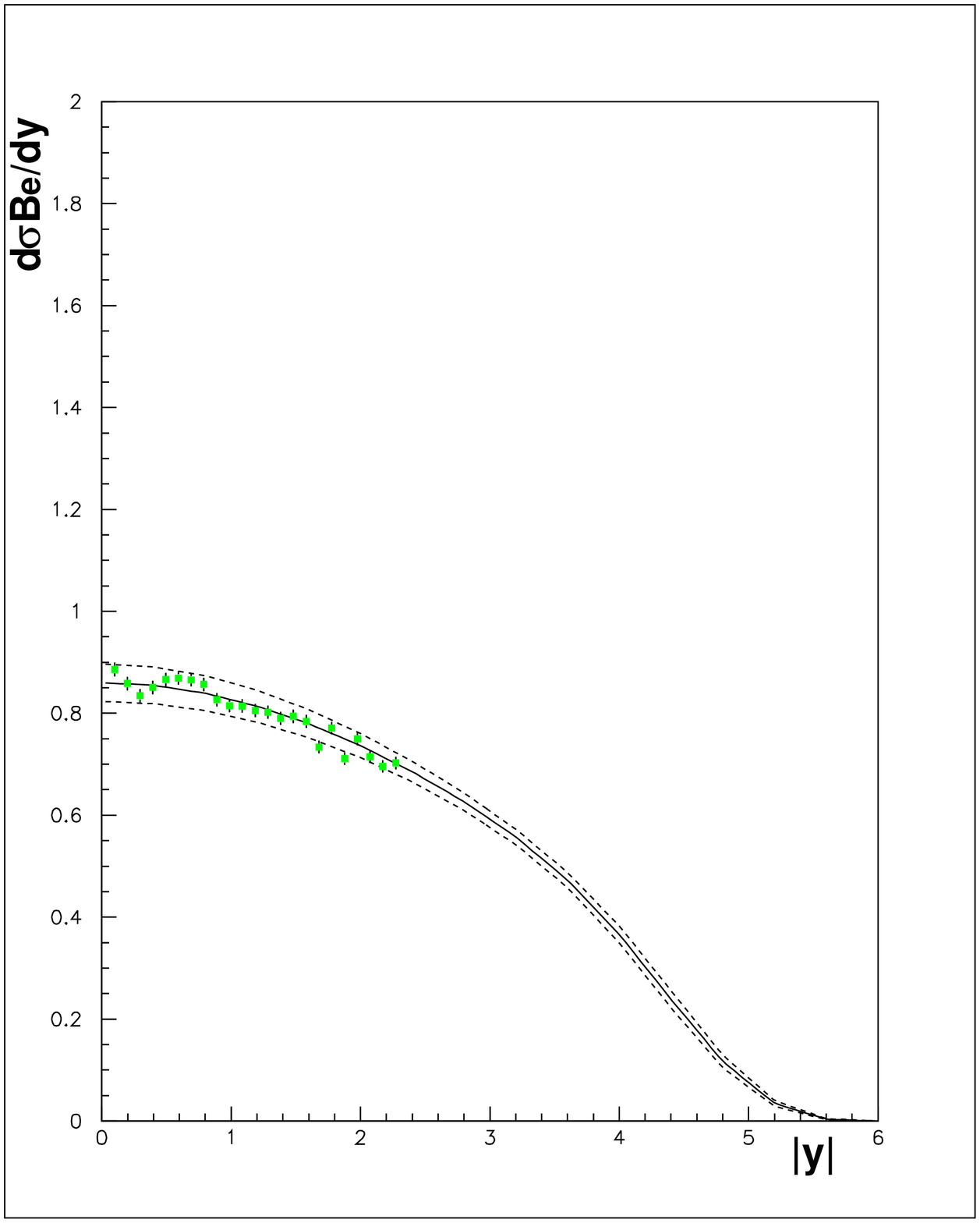,width=0.3\textwidth,height=5cm}
}
\caption {Top row: $e^+$ and $e^-$ rapidity spectra generated from CTEQ6.1 PDFs compared to the analytic prediction
using ZEUS-S PDFs. Bottom row: the same lepton rapidity spectra compared to the analytic 
prediction AFTER including these lepton pseudo-data in the ZEUS-S PDF fit.}
\label{fig:ctqfit}
\end{figure}

\begin{figure}[tbp] 
\vspace{-0.5cm}
\centerline{
\epsfig{figure=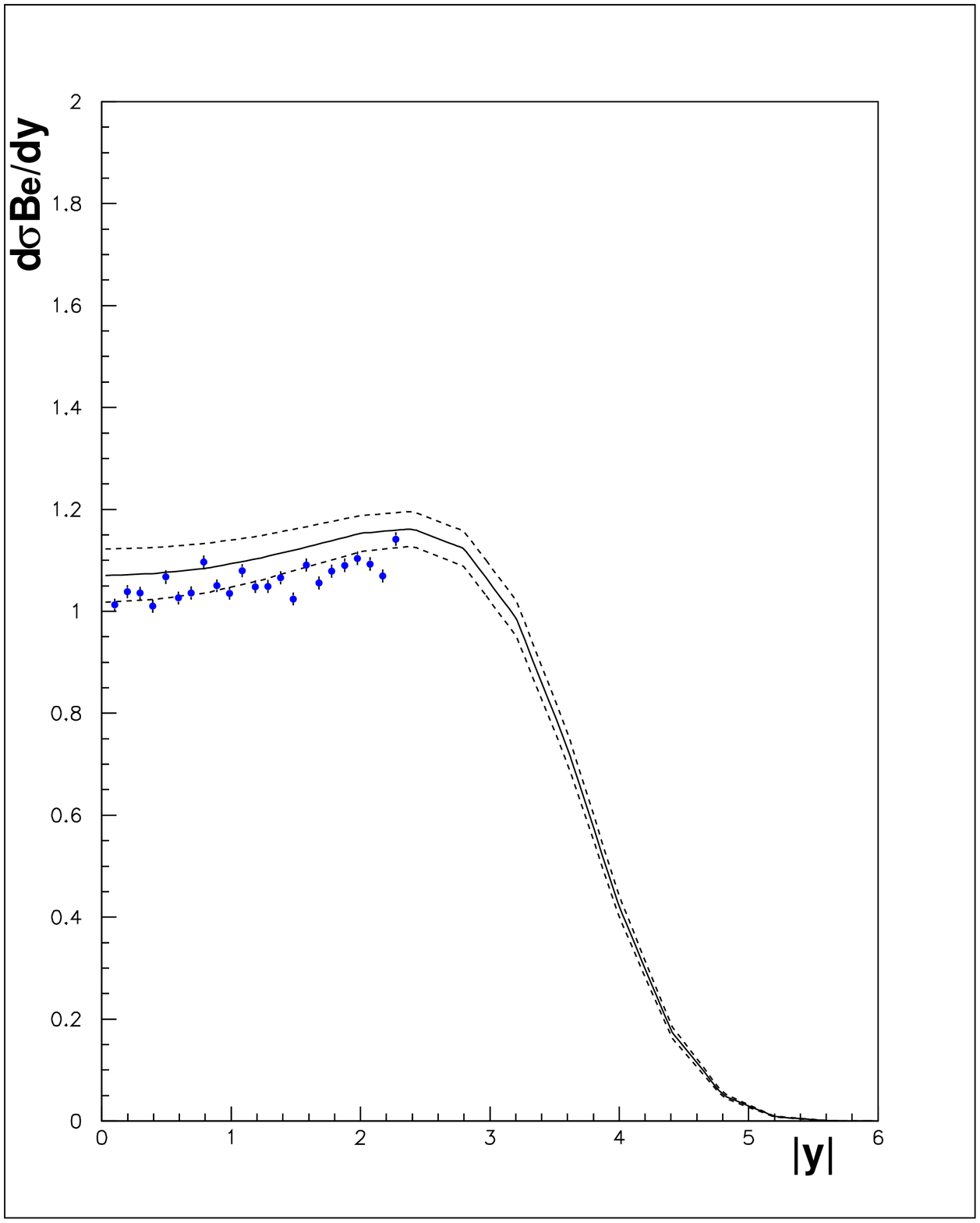,width=0.3\textwidth,height=5cm}
\epsfig{figure=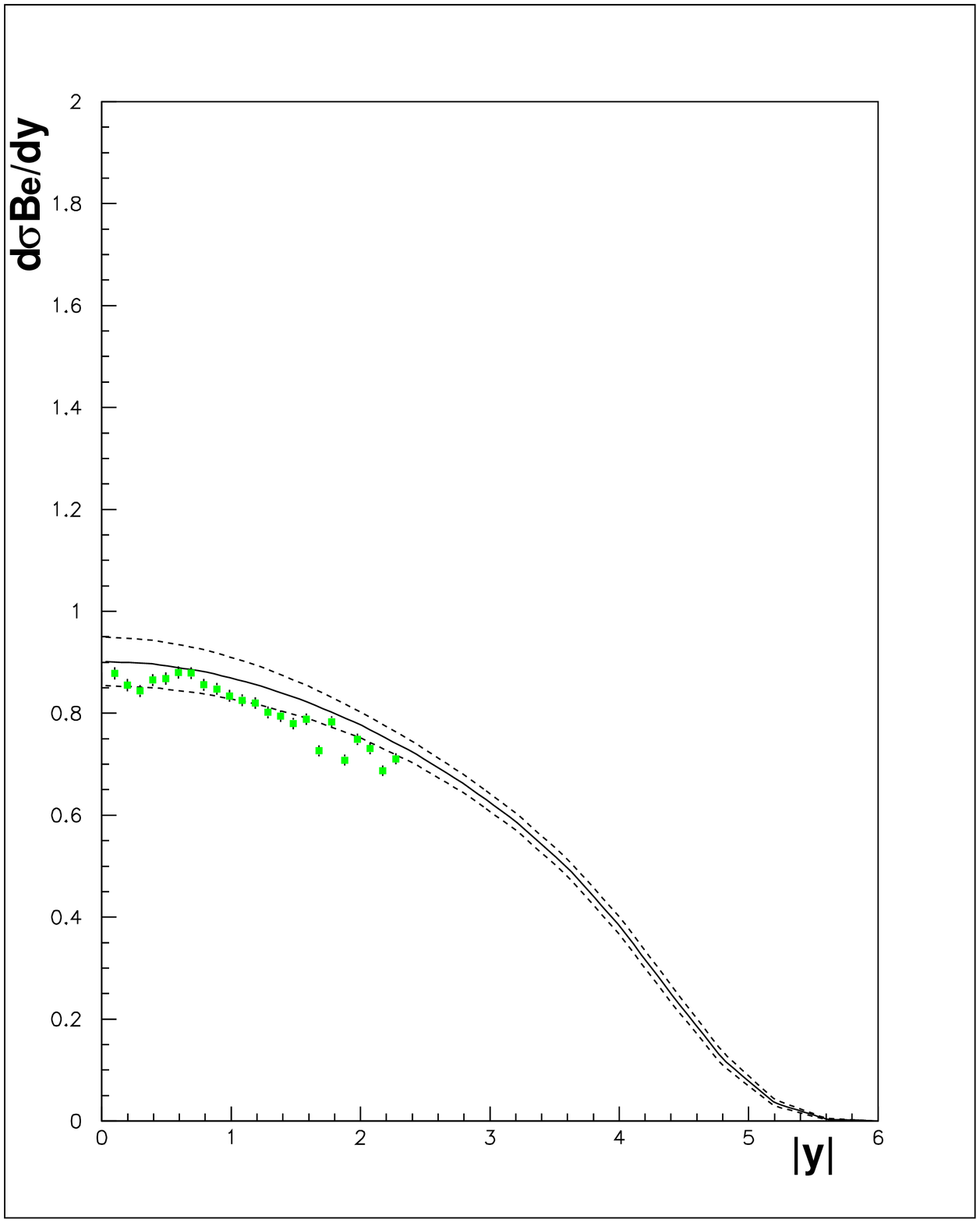,width=0.3\textwidth,height=5cm}
}
\centerline{
\epsfig{figure=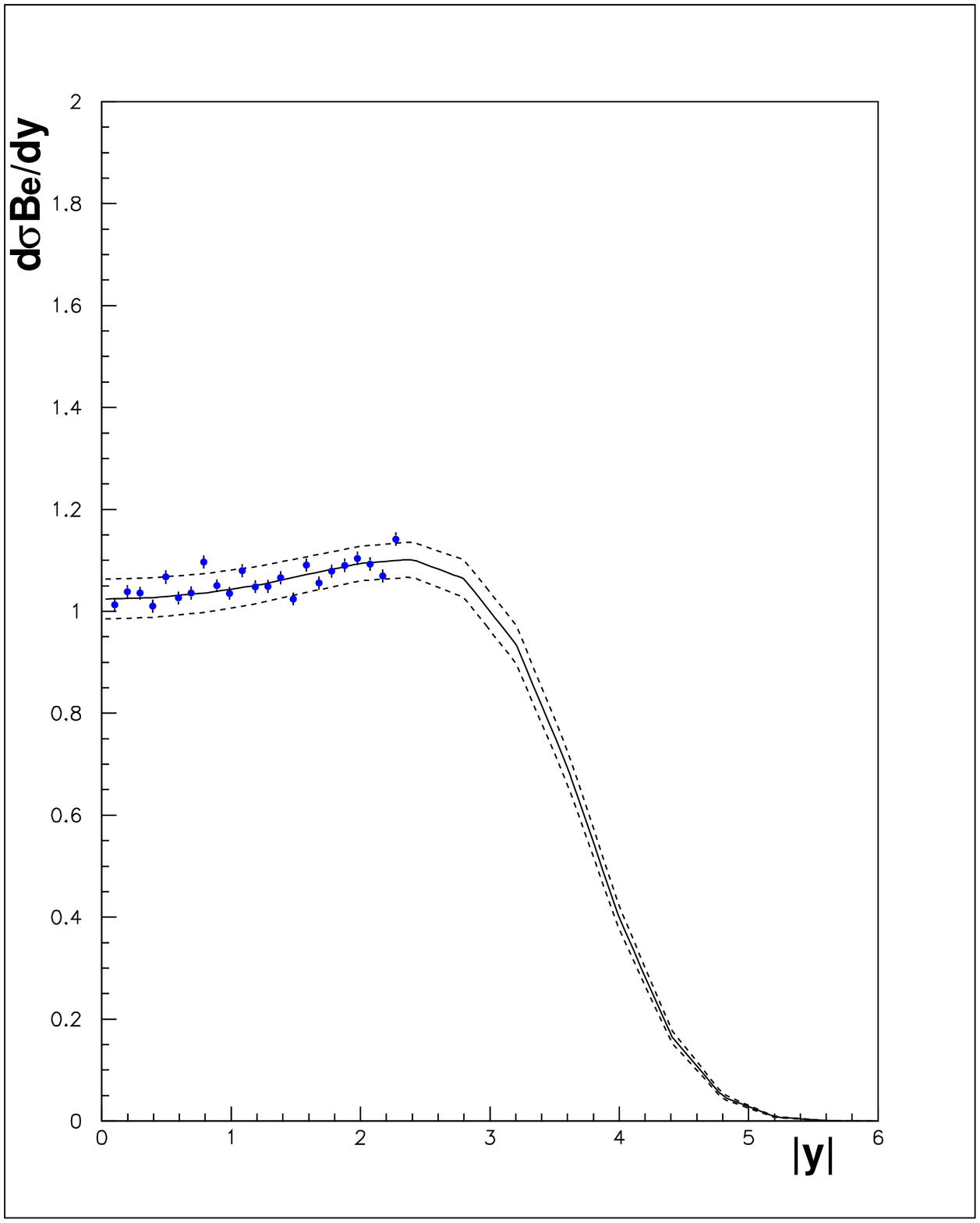,width=0.3\textwidth,height=5cm}
\epsfig{figure=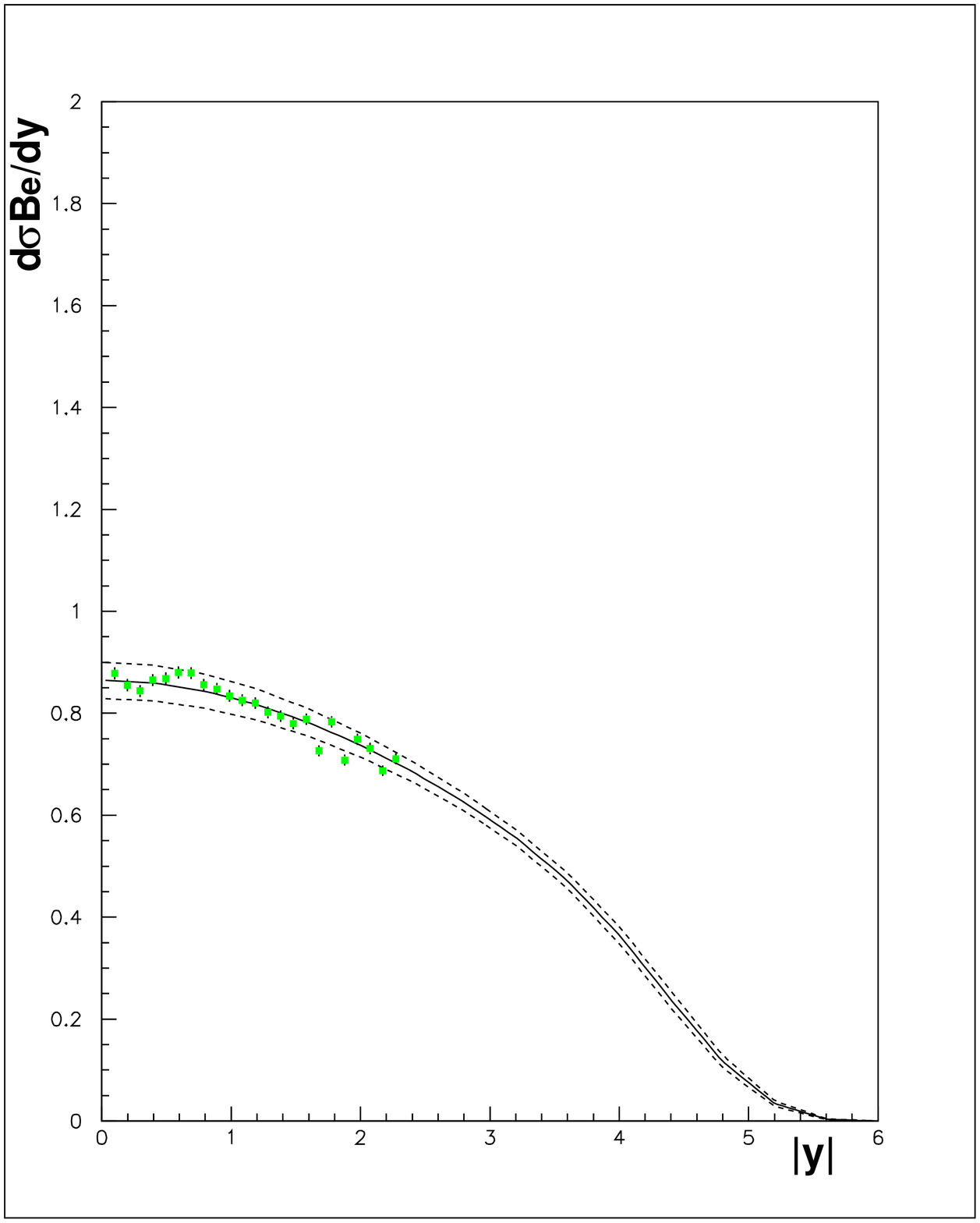,width=0.3\textwidth,height=5cm}
}
\caption {Top row: $e^+$ and $e^-$ rapidity spectra generated from CTEQ6.1 PDFs, which have been passed through 
the ATLFAST detector simulation and corrected back to generator level using ZEUS-S PDFs, 
compared to the analytic prediction
using ZEUS-S PDFs. Bottom row: the same lepton rapidity spectra compared to the analytic 
prediction AFTER including these lepton pseudo-data in the ZEUS-S PDF fit.}
\label{fig:ctqcorfit}
\end{figure}

 The high cross-sections for $W$ production at 
the LHC ensure that it will be the experimental systematic errors, rather than the statistical errors, which 
are determining. We have imposed a random  $4\%$
scatter on our samples of one million $W$ events, generated using different PDFs,
 in order to 
investigate if measurements at this level of precision will improve PDF uncertainties at central rapidity 
significantly if they 
are input to a global PDF fit. Fig.~\ref{fig:zeusfit} shows the $e^+$ and $e^-$ rapidity 
spectra for events generated from the ZEUS-S PDFs ($|\eta| < 2.4$) compared to the analytic 
predictions for these same
 ZEUS-S PDFs. The lower half of this figure illustrates the result if these events are then 
included in the ZEUS-S PDF fit. The size of the PDF uncertainties, at $y=0$, 
decreases from $5.8\%$ to $4.5\%$.  
The largest improvement is in the PDF parameter $\lambda_g$ controlling the 
low-$x$ gluon at the input scale, $Q^2_0$: $xg(x) \sim x^{\lambda_g}$ at low-$x$, $\lambda_g = -0.199 \pm 0.046$, 
before the input of the LHC pseudo-data, compared to, $\lambda_g = -0.196 \pm 0.029$, after input. 
Note that whereas the relative normalisations of the $e^+$ and $e^-$ spectra are set by the PDFs, 
the absolute normalisation of the data is free in the fit so that no assumptions are made on 
our ability to measure luminosity.
Secondly, we repeat this procedure for events generated using the CTEQ6.1 PDFs. 
As shown in Fig.~\ref{fig:ctqfit}, the cross-section for these events is on the lower edge of the uncertainty 
band of the ZEUS-S predictions. If these events are input to the fit the central value shifts and the 
uncertainty decreases. The value of the parameter $\lambda_g$ becomes, $\lambda_g = -0.189 \pm 0.029$, 
after input of these pseudo-data.
Finally to simulate the situation which really faces experimentalists we generate events with CTEQ6.1, 
and pass them through the ATLFAST detector simulation and cuts. We then correct back from detector level 
to generator level using a different PDF set- in this cases the ZEUS-S PDFs- since in practice we will not know 
the true PDFs. Fig.~\ref{fig:ctqcorfit} shows that the resulting corrected data look 
pleasingly like CTEQ6.1, but they are more smeared. When these data are input to the PDF fit the 
central values shift and 
errors decrease just as for the perfect CTEQ6.1 pseudo-data. The value of $\lambda_g$ becomes,
 $\lambda = -0.181 \pm 0.030$, after input of these pseudo-data. Thus we see that the bias introduced by the 
correction procedure from detector to generator level is small compared to the PDF uncertainty. 

\section{Conclusions and a warning: problems with the theoretical predictions at small-$x$?}

We have investigated the PDF uncertainty on the predictions for $W$ and $Z$ production at the LHC, taking 
into account realistic expectations for measurement accuracy and the cuts on data which will be needed to 
identify signal events from background processes. We conclude that at the present level of PDF uncertainty
the decay lepton asymmetry, $A_l$, will be a useful standard model benchmark measurement, and that the 
decay lepton spectra can be used as a luminosity monitor which will be good to $\sim 8\%$. However, 
we have also investigated the measurement accuracy 
necessary for early measurements of these decay lepton spectra to be useful in further constraining the 
PDFs. A systematic measurement error of less than $\sim 4\%$ would provide useful extra constraints.

However, a caveat is that the current study has been performed using 
standard PDF sets which are extracted using NLO QCD in the DGLAP~\cite{ap,*gl,*l,*d} formalism. 
The extension to NNLO is 
straightforward, giving small corrections $\sim 1\%$. PDF analyses at NNLO 
including full accounting of the PDF 
uncertainties are not extensively available yet, so this small correction is not pursued here. However, 
there may be much larger  
uncertainties in the theoretical calculations because the kinematic region involves  
low-$x$.  There may be a need to account for $ln(1/x)$ resummation (first considered in the 
BFKL~\cite{lip,*klf,*bl,*lip2} formalism) 
or high gluon density effects. See reference~\cite{dcs} for a review. 

\begin{figure}[tbp] 
\vspace{-1.0cm}
\centerline{
\epsfig{figure=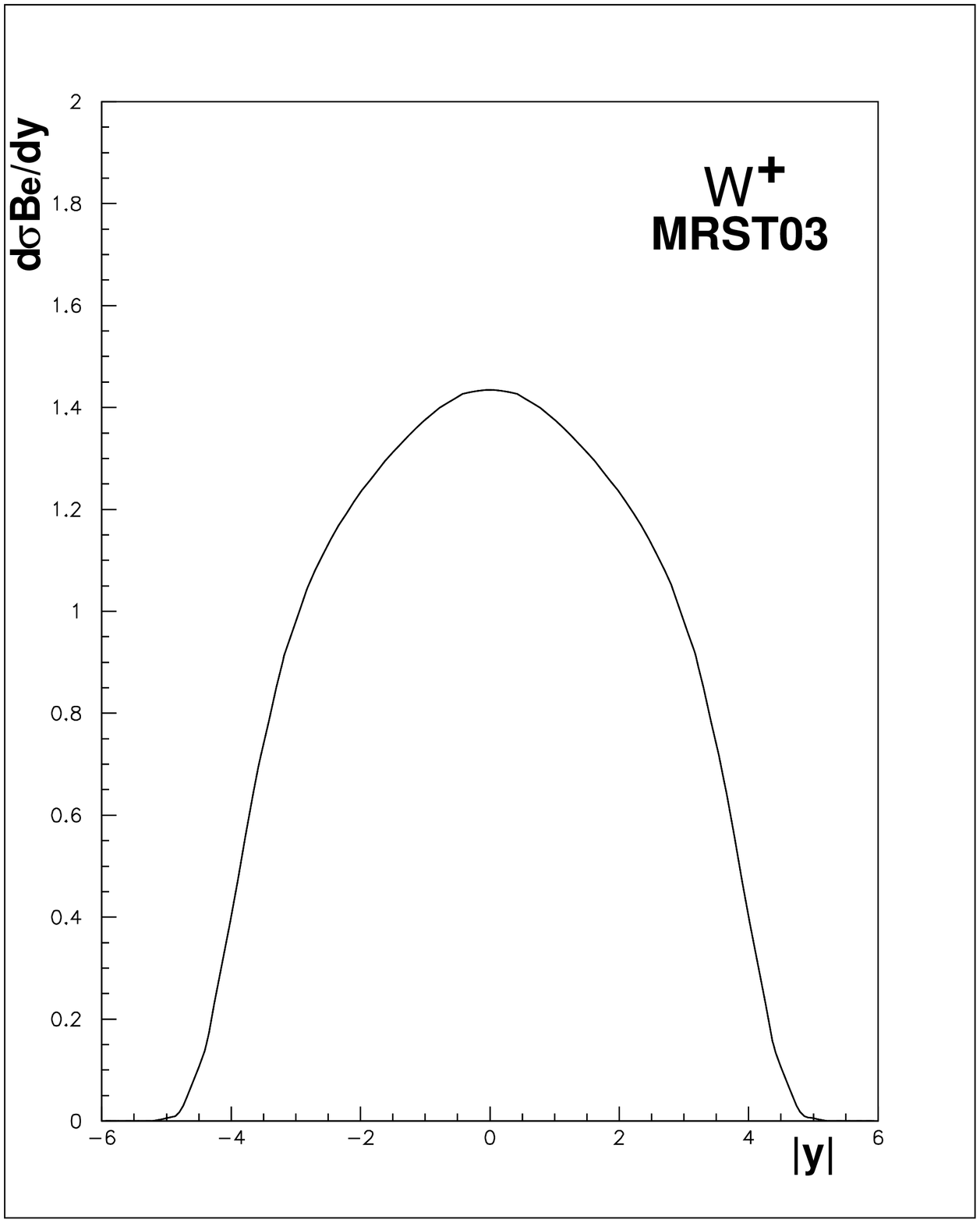,width=0.3\textwidth,height=5cm}
\epsfig{figure=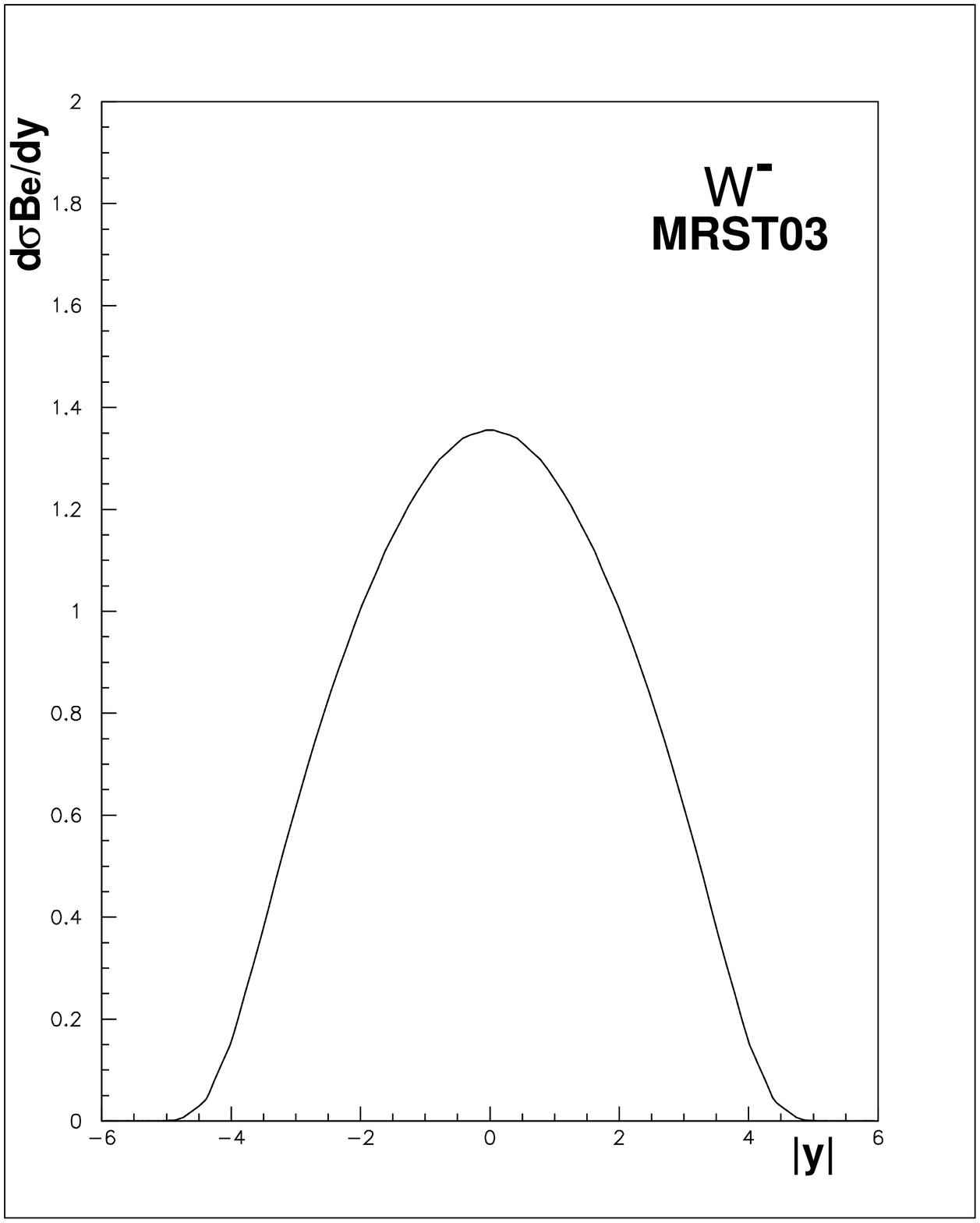,width=0.3\textwidth,height=5cm}
\epsfig{figure=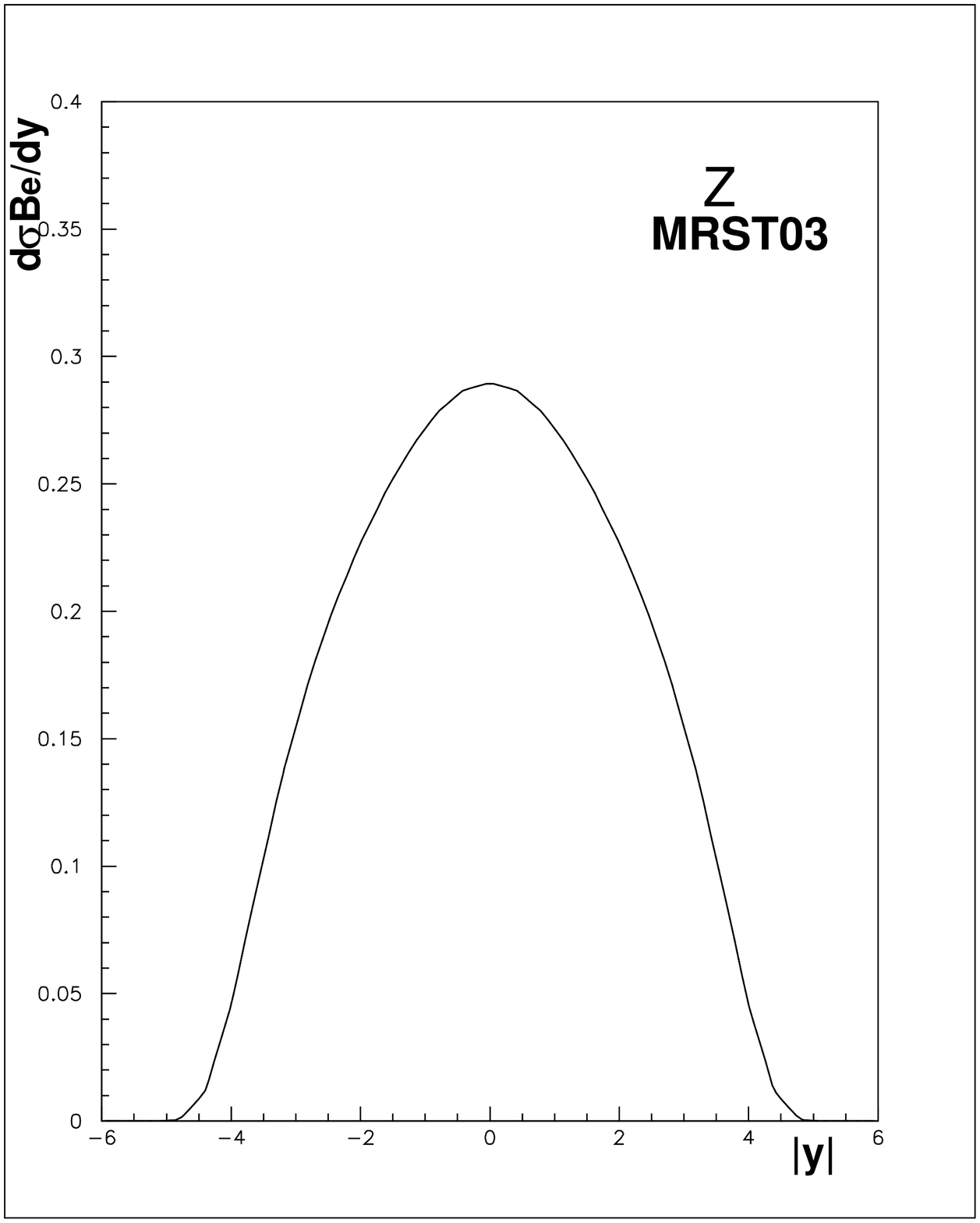,width=0.3\textwidth,height=5cm} 
}
\caption {LHC $W^+,W^-,Z$ rapidity distributions for the MRST03 PDFs: left plot $W^+$; middle plot $W^-$; 
right plot $Z$}
\label{fig:mrst03pred}
\end{figure}

The MRST group recently produced a PDF set, MRST03, which does not include any data for $x < 5\times 10^{-3}$.
The motivation behind this was as follows. In a global DGLAP fit to many data sets there is always a certain 
amount of tension between data sets. This may derive from the use of an inappropriate theoretical formalism
for the kinematic range of some of the data. Investigating the effect of kinematic cuts on the data, MRST 
found that a cut, $ x > 5\times 10^{-3}$, considerably reduced tension between the remaining data sets. An 
explanation may be the inappropriate use of the DGLAP formalism at small-$x$. 
The MRST03 PDF set is thus free of this bias 
BUT it is also only valid to use it for $x > 5\times 10^{-3}$. 
What is needed is an alternative theoretical formalism
for smaller $x$. However, the MRST03 PDF set may be used as a toy PDF set, to illustrate the effect of using 
very different PDF sets on our predictions. A comparison of Fig.~\ref{fig:mrst03pred} with 
Fig.~\ref{fig:WZrapFTZS13} or Fig.~\ref{fig:mrstcteq} shows how different the analytic 
predictions are from the conventional ones, and thus illustrates where we might  expect to see differences due 
to the need for an alternative formalism at small-$x$. 

\bibliographystyle{heralhc} 
{\raggedright
\bibliography{heralhc}
}
\end{document}